\documentclass[fleqn,10pt]{wlscirep}
\usepackage[utf8]{inputenc}
\usepackage[T1]{fontenc}
\usepackage{booktabs}
\usepackage{multirow}
\usepackage{enumitem}
\usepackage{subfigure}
\usepackage{lineno}
\title{Understanding occupants' behaviour, engagement, emotion, and comfort indoors with heterogeneous sensors and wearables}

\author[1]{Nan Gao}
\author[2]{Max Marschall}
\author[3]{Jane Burry}
\author[4]{Simon Watkins}
\author[5]{Flora D. Salim}
\affil[1]{RMIT University, School of Computing Technologies, Melbourne, 3000, Australia}
\affil[2]{RMIT University, School of Architecture and Urban Design, Melbourne, 3000, Australia}
\affil[3]{Swinburne University of Technology, School of Design, Melbourne, 3122, Australia}
\affil[4]{RMIT University, School of Engineering, Melbourne, 3000, Australia}
\affil[5]{University of New South Wales (UNSW), School of Computer Science and Engineering, Sydney, 1466, Australia}

\begin{abstract}
We conducted a field study at a K-12 private school in the suburbs of Melbourne, Australia. The data capture contained two elements: First, a 5-month longitudinal field study \textit{In-Gauge} using two outdoor weather stations, as well as indoor weather stations in 17 classrooms and temperature sensors on the vents of occupant-controlled room air-conditioners; these were collated into individual datasets for each classroom at a 5-minute logging frequency, including additional data on occupant presence. The dataset was used to derive predictive models of how occupants operate room air-conditioning units. Second, we tracked 23 students and 6 teachers in a 4-week cross-sectional study \textit{En-Gage}, using wearable sensors to log physiological data, as well as daily surveys to query the occupants' thermal comfort, learning engagement, emotions and seating behaviours. Overall, the combined dataset could be used to analyse the relationships between indoor/outdoor climates and students' behaviours/mental states on campus, which provide opportunities for the future design of intelligent feedback systems to benefit both students and staff.

\end{abstract}
\begin{document}

\flushbottom
\maketitle

\thispagestyle{empty}

\section*{Background \& Summary}
How can indoor spaces be designed in ways that increase occupant well-being while decreasing energy consumption? Answering this question requires a holistic understanding of indoor climates, occupant comfort and behaviour, as well as the dynamic relationships between these different aspects. The present study sits within a context of research that aims to gain insights by examining these themes using mixed methods of data capture within operational buildings. More specifically, the study contains two separate assays, each relating to a distinct body of existing research.

The first assay is a 5-month longitudinal field study using outdoor and indoor weather stations as well as sensors to determine the use of occupant-controlled room air-conditioners. This assay was undertaken to contribute knowledge to the research field of occupant behaviour modelling in building performance simulation. During the design of buildings, engineers often use simulations to predict the indoor environmental quality and energy consumption of design options in order to inform decision-making. There are often large discrepancies between simulated and actual building performance \cite{haldi2011impact}. One of the main factors driving this so-called 'performance gap' is the current misrepresentation of occupant behaviour in the simulations \cite{rijal2011algorithm}. The software is accurate at modelling deterministic systems like automated air-conditioning units that are governed by set point temperatures, but incapable of accurately modelling the probabilistic nature of human behaviour, for example, the manual operation of air-conditioners. Occupant behaviour tends to be modelled on simplistic, rule-of-thumb assumptions that are not backed by data \cite{schiavon2013dynamic}, usually by using the same set point approaches that are applied to automated systems (e.g. occupant switches on the air-conditioner when the indoor temperature exceeds 24°C). Actual human behaviour is less responsive and more varied; thus, researchers have conducted field studies in operational buildings, by measuring various environmental and other variables alongside an observed behaviour (for example, the operation of air-conditioners, windows, lights, fans, etc.). They use this data to derive statistical models of the observed behaviour based on one or several of the observed independent variables  \cite{schweiker2012verification,langevin2015tracking,reinhart2003monitoring}. The first assay of our study contributes data towards this endeavour, specifically enabling the creation of predictive models of occupants’ use of room air-conditioners in schools.

The second assay is a four-week cross-sectional study tracking 23 students and six teachers, using wearable sensors to log physiological data, as well as daily surveys to query the occupants' thermal comfort, learning engagement, seating locations and emotions while at school. Buildings contribute about a third of world energy consumption, mainly due to the use of heating, ventilation and air-conditioning (HVAC) systems for indoor climate regulations. Considering that we have spent so much energy and effort in providing adequate environments for building occupants, it is worth investigating what exactly constitutes their comfort and well-being. The above-mentioned building performance simulations tend to define comfort either by using deemed-to-satisfy temperature thresholds or by using comfort models, most commonly the predicted mean vote (PMV) model. However, the PMV model has not been updated since it was derived from laboratory experiments in the 1960s. It has been criticised for its poor predictive performance in real-world contexts \cite{cheung2019analysis} and does not appear to apply to all age groups \cite{kim2018thermal}. Furthermore, thermal acceptance is clearly only one of several metrics for assessing indoor well-being. 

\begin{table}[]
\centering
\footnotesize
\setlength{\tabcolsep}{5pt}
\renewcommand{\arraystretch}{1.1}
\begin{tabular}{@{}|l|l|l|l|l|l|l|l|@{}}
\hline
Name  & Year & Par. & Type & Modalities                                                                      & Annotations                                                                                                         & Duration                                                       & Scenario                                                                                         \\ \hline
Driving-stress \cite{data_drivingstress} & 2005          & 24           & Field         & ECG, EDA, EMG, RESP                                                                       & Stress level                                                                                                                 & > 50 minutes                                                    & Real-world driving tasks                                                                                  \\ \hline
DEAP  \cite{data_deap}         & 2011          & 32           & Lab           & \begin{tabular}[c]{@{}l@{}}Videos, EEG, EDA, BVP,\\ RESP, ST, EMG and EOG\end{tabular}    & \begin{tabular}[c]{@{}l@{}}Arousal, valence, like/\\ dislike, dominance, \\ familarity\end{tabular}                          & 40 minutes                                                                 & Watch music videos                                                                                        \\ \hline
Driving-work \cite{data_drivingwork}   & 2013          & 10           & Field         & EDA, HR, TEMP                                                                             & Mental workload                                                                                                              & 30 minutes                                                                 & Drive a predefined route                                                                                  \\ \hline
StudentLife \cite{data_studentlife}   & 2014          & 48           & Field         & Smartphone                                                                                & Stress, mood, happiness                                                                                                      & 10 weeks                                                                & Real life, student exams                                                                                  \\ \hline
DECAF   \cite{data_decaf}       & 2015          & 30           & Lab           & \begin{tabular}[c]{@{}l@{}}ECG, EMG, EOG, MEG,\\ near-infrared face, video\end{tabular}   & \begin{tabular}[c]{@{}l@{}}Valence, arousal, and \\ dominance\end{tabular}                                                   & > 1 hour                                                     & \begin{tabular}[c]{@{}l@{}}Watch music video and \\ movie clips\end{tabular}                              \\ \hline
Non-EEG  \cite{data_noneeg}      & 2016          & 20           & Lab           & \begin{tabular}[c]{@{}l@{}}ACC, EDA, HR, TEMP, \\ SpO2\end{tabular}                       & N/A                                                                                                                          & < 1 hour                                                       & \begin{tabular}[c]{@{}l@{}}Four types of stress \\ (physical, emotional, \\ cognitive, none)\end{tabular} \\ \hline
Ascertain  \cite{data_ascertain}    & 2016          & 58           & Lab           & \begin{tabular}[c]{@{}l@{}}ECG, EDA, EEG, facial\\ features\end{tabular}                  & \begin{tabular}[c]{@{}l@{}}Arousal, valence, \\ engagement, liking, \\ familarity, personality\end{tabular}                  & 90 minutes                                                                 & Watch movie clips                                                                                         \\ \hline
Stress-math  \cite{data_stressmath}  & 2017          & 21           & Lab           & ACC, EDA, HR, TEMP                                                                        & Anxiety                                                                                                                      & \begin{tabular}[c]{@{}l@{}}26 hours\\ (total)\end{tabular} & \begin{tabular}[c]{@{}l@{}}Solve math questions \\ under different pressure\end{tabular}                  \\ \hline
WESAD  \cite{data_wesad}        & 2018          & 15           & Lab           & \begin{tabular}[c]{@{}l@{}}ACC, BVP, ECG, EDA, \\ EMG, RESP, TEMP\end{tabular}            & Affect, anxiety, stress                                                                                                      & 2 hours                                                                 & \begin{tabular}[c]{@{}l@{}}Neutral, amusement and\\ stress conditions\end{tabular}                        \\ \hline
Snake    \cite{data_cogload_snake}      & 2020          & 23           & Lab           & ACC, BVP, EDA, TEMP                                                                       & \begin{tabular}[c]{@{}l@{}}Cognitive load, \\ personality\end{tabular}                                                       & > 6 minutes                                                     & \begin{tabular}[c]{@{}l@{}}Smartphone games with\\ three difficulty levels\end{tabular}                   \\ \hline
CogLoad    \cite{data_cogload_snake}    & 2020          & 23           & Lab           & ACC, BVP, EDA, TEMP                                                                       & \begin{tabular}[c]{@{}l@{}}Cognitive load, \\ personality\end{tabular}                                                       & N/A                                                                     & 6 cognition load tasks                                                                                    \\ \hline
K-EmoCon \cite{data_kemocon}      & 2020          & 32           & Lab           & \begin{tabular}[c]{@{}l@{}}Videos, audio, ACC, \\ EDA, EEG, ECG, BVP,\\ TEMP\end{tabular} & \begin{tabular}[c]{@{}l@{}}Arousal, valence,\\ stress, affect\end{tabular}                                                   & \begin{tabular}[c]{@{}l@{}}173 minutes\\ (total)\end{tabular}           & \begin{tabular}[c]{@{}l@{}}Social interaction \\ scenario involving \\ two people\end{tabular}            \\ \hline
En-Gage         & 2022          & 29           & Field         & \begin{tabular}[c]{@{}l@{}}ACC, EDA, BVP, TEMP,\\ In. TEMP, HUMID., CO2,\\ NOISE\end{tabular}
& \begin{tabular}[c]{@{}l@{}}Cognitive, behavioural,\\ emotion engagement, \\ thermal comfort,  arousal,\\ valence\end{tabular} & \begin{tabular}[c]{@{}l@{}}4 weeks\\ (1416 hours\\ in total)\end{tabular}                     & \begin{tabular}[c]{@{}l@{}}Real-world courses in\\  a high school\end{tabular}                            \\  \hline
\end{tabular}
\caption{Publicly available datasets in the affective computing area.}
\label{tab: datasets}
\end{table}

On the other hand, studying student engagement, emotions, and daily behaviours has attracted increasing interest to address problems such as low academic performance and disaffection. Sensor-based physiological and behaviour recordings provide great opportunities to unobtrusively measure students' behaviours and emotional changes in classroom settings \cite{gao2020n,di2018unobtrusive}. In previous studies, various physiological signals, such as  electrodermal activity (EDA) and heart rate variability (HRV), and environmental data have been explored to assess emotional arousal and engagement levels. For example, EDA is generally considered to be a good indicator of psychological arousal and has been increasingly studied for the detection of engagement \cite{gao2020n,di2018unobtrusive}, emotion \cite{bakker2011s}, and depression \cite{sarchiapone2018association}, etc. Existing datasets in affective computing either provide limited scope for understanding emotion responses in real-world settings or only consider a particular type of annotation to meet their research goals (e.g., stress level and mental workload). Table~\ref{tab: datasets} shows how \textit{En-Gage} dataset is distinguished from existing emotion datasets. 

\textit{En-Gage} is the first publicly available dataset  studying the daily behaviours and engagement of high school students using heterogeneous sensing. Together with \textit{In-Gauge} dataset, it offers a unique opportunity to analyse the relationships between indoor climates and the mental states of school students - not only related to their thermal comfort but also their emotions, engagement and productivity while at school. Especially, it's unusual to combine individual sensor data with building environmental data together, to study how indoor and outdoor environments influence the complex occupant behaviours and physiological responses. The combination of two datasets will benefit building scientists, behaviour psychologists and affective computing researchers in future research. 

\section*{Methods}

\subsection*{Ethics approval}
The data collection was approved by the Science, Engineering and Health College Human Ethics Advisory Network (SEH CHEAN) of RMIT University. SEH CHEAN also reviewed and approved the consent forms for participants and guardians of minors, which included information on the purpose of and procedures for the research, the types of data to be collected, the compensation for the involvement and the protocols for privacy protection and data storage. The project was also approved by the principal of the school in which the study was conducted.

\begin{table}
\small
\centering
\renewcommand{\arraystretch}{1.1}
\begin{tabular}{|l|l|l|}
\hline
Group                     & Room & Participant                                                 \\ \hline
\multirow{3}{*}{Form}     & R1   & P13, P14, P15, P16, P17, P18, P19, P20, P21, P22            \\
                          & R2   & P8, P9, P10, P11, P12, P23                                  \\
                          & R3   & P1, P2, P3, P4, P5, P6, P7                                  \\\hline
\multirow{3}{*}{Maths}     & R1   & P2, P4, P5, P10, P11, P14, P18                              \\
                          & R2   & P3, P6, P7, P8, P9, P15, P16, P17, P20                      \\
                          & R3   & P1, P12, P13, P19, P21, P22, P23                            \\\hline
\multirow{4}{*}{Language} & R1   & P1, P2, P4, P7, P10, P13, P15, P17, P19, P20, P21, P22, P23 \\
                          & R2   & P9, P14                                                     \\
                          & R3   & P5, P6, P11, P12, P16                                       \\
                          & R4   & P3, P8 P18                                                  \\ \hline
\end{tabular}
\caption{Distribution of student participants in different class groups.}
\label{tab:classgroup}
\end{table}

\begin{table}
\centering
\small
\renewcommand{\arraystretch}{1.1}
\begin{tabular}{|l|l|l|l|}
\hline Devices
                                                                                        & Collected data                                                                                                                           & Sampling rate & Time frame                                                                    \\ \hline
\multirow{4}{*}{Empatica E4 wristband}                                                  & 3-axis acceleration                                                                                                                        & 32 Hz         & \multirow{4}{*}{\begin{tabular}[c]{@{}l@{}}4 weeks \\ \end{tabular}} \\
                                                                                        & Skin temperature                                                                                                                           & 4 Hz          &                                                                                  \\
                                                                                        & Electrodermal activity                                                                                                                     & 4 Hz          &                                                                                  \\
                                                                                        & Blood volume pulse                                                                                                                         & 64 Hz         &                                                                                  \\ \hline
\begin{tabular}[c]{@{}l@{}} Netatmo indoor weather\\station\end{tabular}                 & Humidity, temperature, noise level, CO$_2$                                                                                                          & 5 minutes      & 5.5 months                                                                         \\\hline
\begin{tabular}[c]{@{}l@{}} DigiTech XC0422 outdoor\\ weather station\end{tabular}               & \begin{tabular}[c]{@{}l@{}}Temperature, humidity, barometric \\ pressure, wind speed, wind direction, \\ solar radiation, UV, rainfall\end{tabular} & 5 minutes      & 5.5 months                                                                         \\\hline
\begin{tabular}[c]{@{}l@{}} PHILIO Z-wave (attached \\ to air-conditioning vents)\end{tabular} & Humidity, temperature                                                                                                                      & 5 minutes           & 5.5 months                                                                         \\ \hline
\end{tabular}
\caption{Data collected with sensors with respective sampling rate and time.}
\label{tab:instrument}
\end{table}
\subsection*{Participants and recruitment}
For the cross-sectional study, we recruited participants from a K-12 private school in the suburbs of Melbourne (pop. 700). The recruitment occurred  in August and September 2019, and calls for participation were disseminated through information leaflets, recruitment letters and a presentation in the school hall, with the assistance of the director teacher of the Year 10 students (Year 10 is the eleventh year of compulsory education in Australia). The admission was restricted to Year 10 students and their teachers whose native language was English or who were bilingual. A total of 23 (15--17 years old, 13 female and 10 male) out of 75 Year 10 students and six (33-62 years old, four female and two male) out of 12 teachers met the inclusion criteria, volunteered for the study and signed the consent forms. Since all the student participants were underage, their guardians also provided signed consent forms. Raw data for \textit{n} = 23 student participants were properly recorded and nearly complete (but with different wristband wearing days), constituting the majority of the \textit{En-Gage} dataset.

The volunteers were then asked to complete an online background survey, which was accessible through a web page link that was shared with them. In the survey, we collected information on the participants' age, gender, general thermal comfort and classes. The Year 10 students at the school were taught in separate class groups. They were separated into three \textit{Form} groups for English, Science, Global Politics, Physical Education and Health/Sport courses, three \textit{Maths} groups and four \textit{Language} groups (see Table \ref{tab:classgroup}). Asking for each student's class group in the background survey allowed us to determine which classroom they were in at any given time. Among the participating teachers, there were three math teachers, one English teacher, one Japanese teacher, and one science teacher.

As a token of appreciation for their participation, we awarded each participating student with a certificate of participation and four movie vouchers - one for each week of successful participation. Participation in this research project was voluntary, and we communicated to participants that they were free to withdraw from the project at any stage.

\subsection*{Experiment setup}
We conducted our study at a mixed-gender K-12 private school.
The longitudinal study was conducted for a 5.5-month period from 7 October 2019 to 23 March 2020, using the indoor and outdoor weather stations as well as temperature sensors attached to air-conditioning outlets. The cross-sectional study included four weeks of data capture: the first two weeks of data collection started from 2 September 2019, and the second two weeks of data collection started from 28 October 2019, using the wearable sensors as well as the same weather stations in the first longitudinal study. The two studies are located on the same campus, and the timelines of the two studies were partly overlapped. As a result, the collected data (i.e., weather information and occupant behaviours) in the longitudinal study can benefit the cross-sectional study or vice versa. For instance, the outdoor temperature and humidity can help researchers understand student clothing insulation and thermal comfort on campus. Additionally, the physiological signals can be combined with the environmental signals to accurately predict the heating/cooling behaviours of occupants in buildings.

\begin{figure}
	\centering
	\subfigure[Empatica E4 wristbands \label{fig: 4band}]{\includegraphics[width=0.3\textwidth]{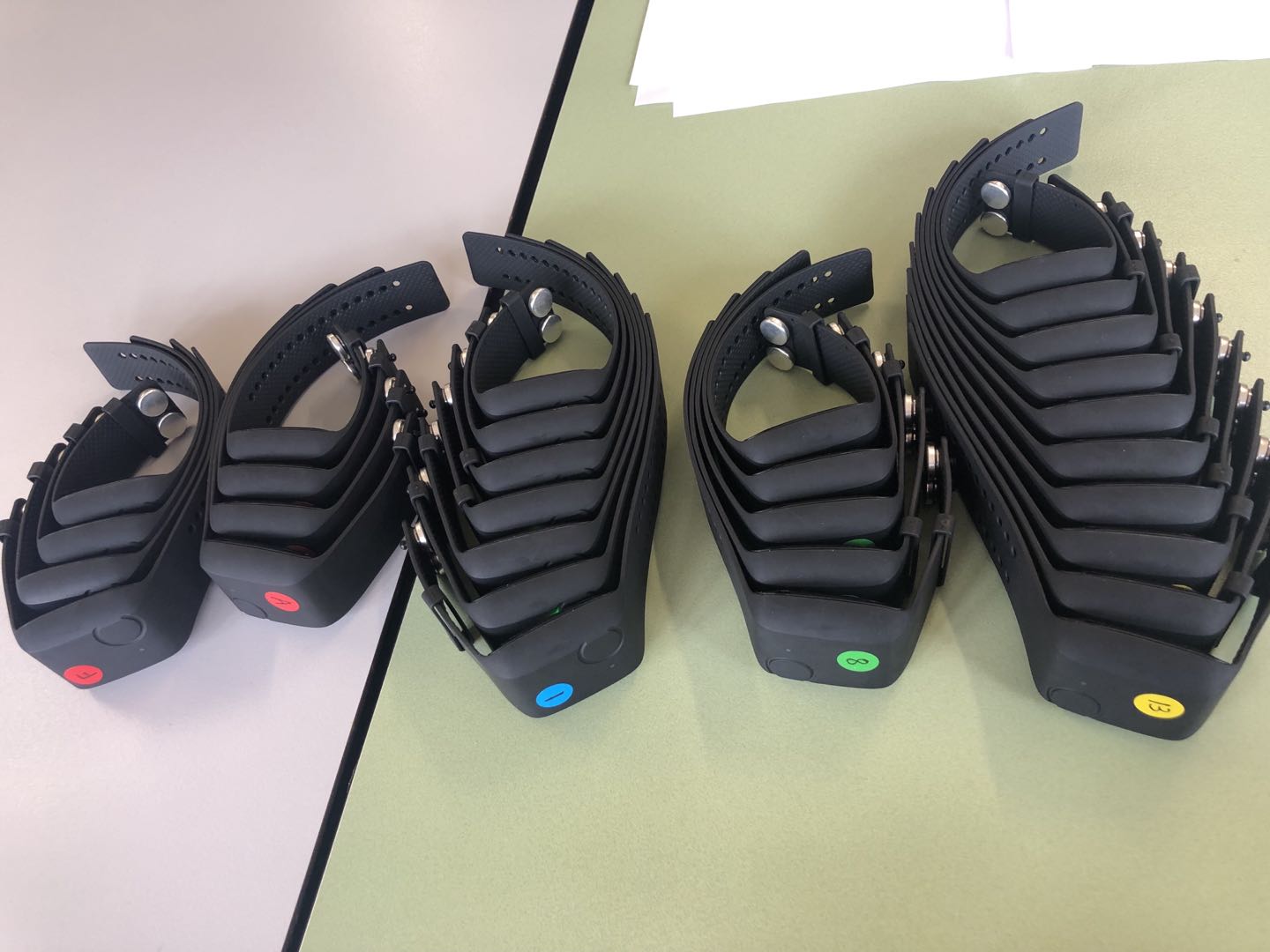}
	\label{fig: sub e4band}}
	\hspace{0.2cm}
	\subfigure[Netatmo indoor weather station]{\includegraphics[width=0.27\textwidth]{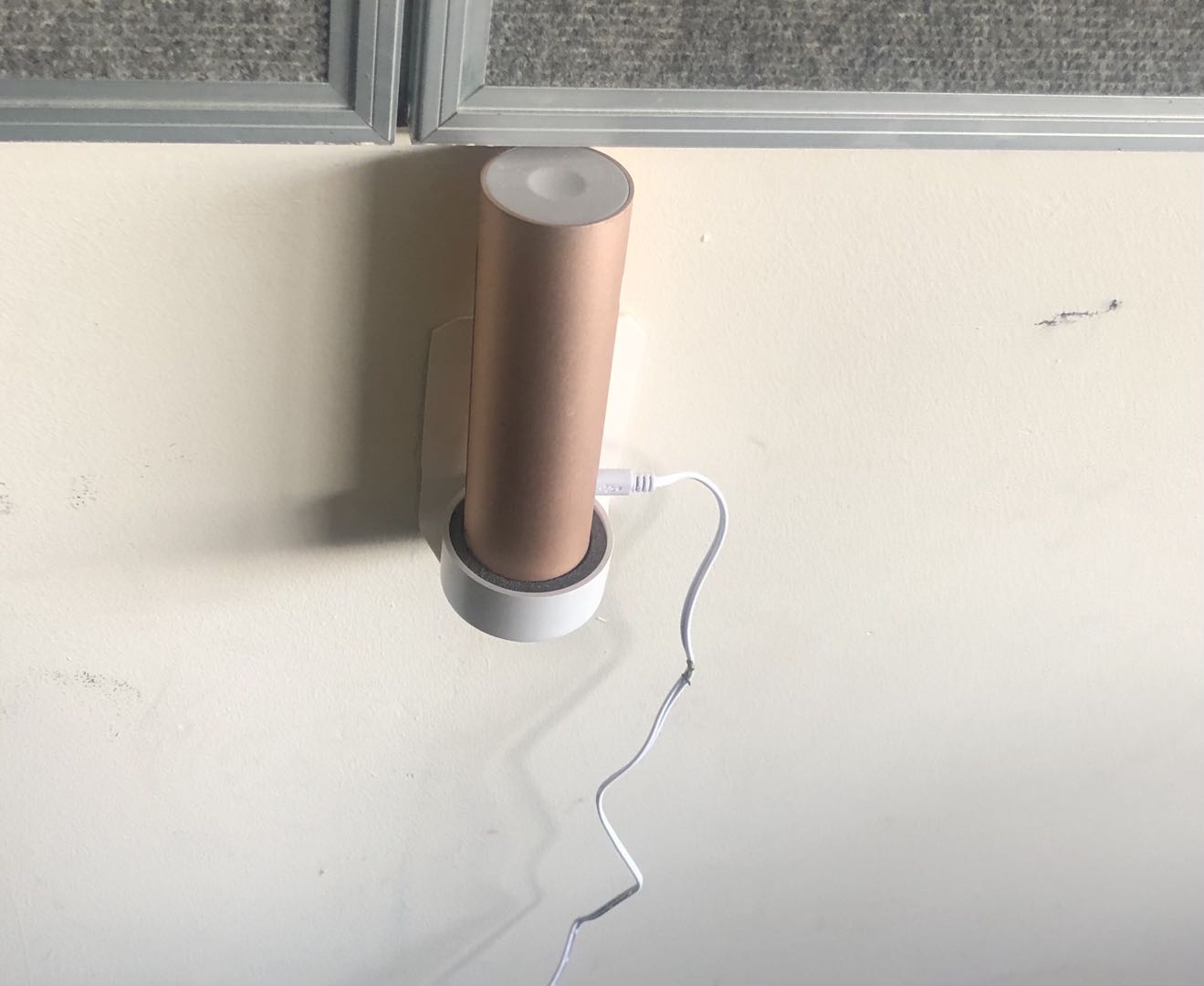}
    \label{fig: sub netamo}}
    \hspace{0.2cm}
    \subfigure[Classroom for Year 10 students]{\includegraphics[width=0.295\textwidth]{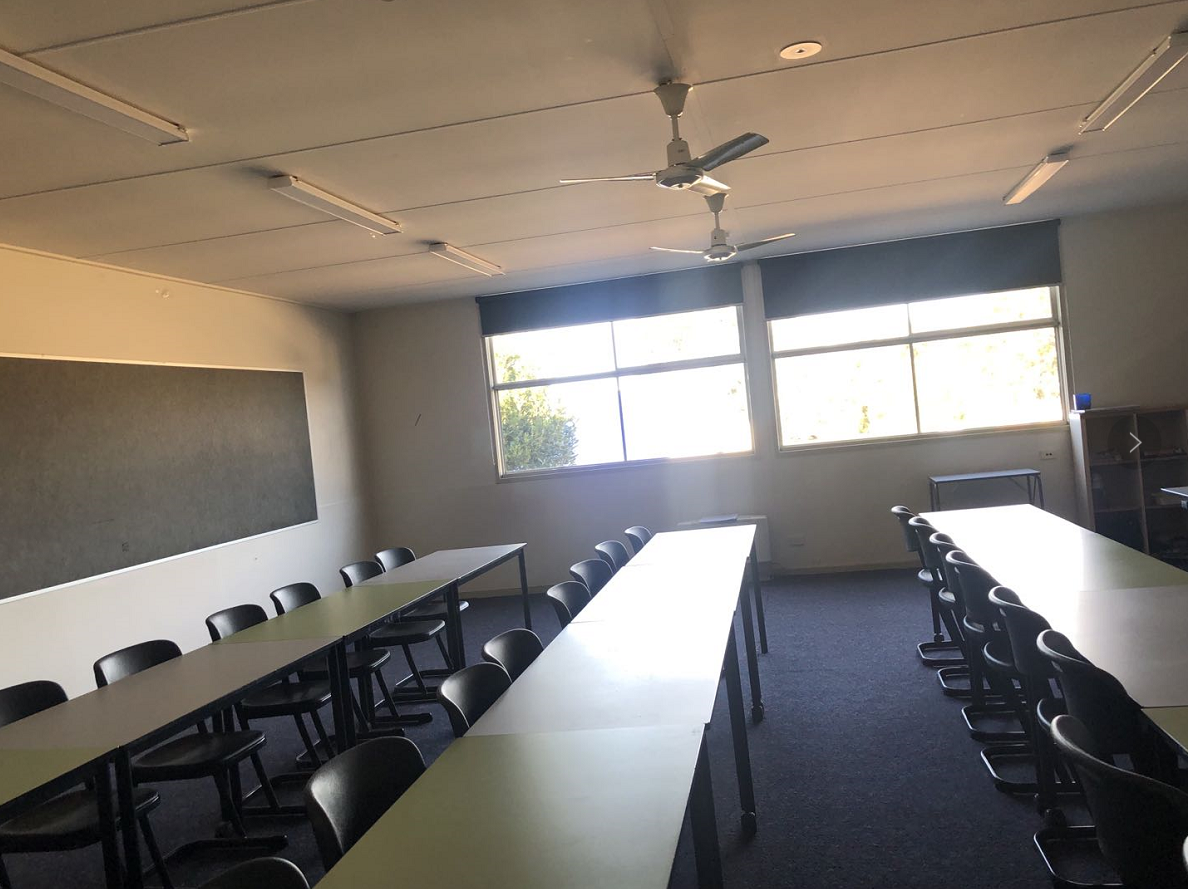}   
    \label{fig: sub room}}
    \caption{Devices and environments for collecting wearable and indoor data.}
    \label{figb: netamo and e4band}
\end{figure}

In the study, we tracked participants using \textit{Empatica E4} wristbands (see Figure \ref{fig: 4band}) to measure physiological data, as well as daily surveys to query their thermal comfort, learning engagement and emotions while at school. Overall, we have collected 488 survey responses and 1415.56 hours of wearable data from all participants. During the data collection, one representative student was selected in each of the three \textit{Form} classes. Their job was to distribute wristband sensors each morning, collect them after school and remind participants to complete the online surveys at the appropriate times. We anonymised the student's data by assigning each student an identity number (ID). Occupancy schedules were obtained from the individual classroom schedules provided by the school. These schedules can be used to represent the actual occupancy patterns of the building, although slight deviations from the planned schedule are to be expected in a school setting due to sickness and other circumstances. The following is a description of the research instruments used in the study (see Table~\ref{tab:instrument}).

\begin{table}[]
\centering
\small
\renewcommand{\arraystretch}{1.1}
\begin{tabular}{|l|l|l|}
\hline
Annotation categories                                                                 & Description                                                                                                           & Measurement scale                                                                                                                                              \\ \hline
Thermal sensation                                                                     & Commonly used ASHRAE  thermal sensation \cite{handbook2009american}                                                                                      & \begin{tabular}[c]{@{}l@{}}-3: cold, -2: cool, -1: slightly cool, 0: neutral, \\ 1: slightly warm, 2: warm, 3: hot\end{tabular}                               \\ \hline
Thermal preference                                                                    & Commonly used ASHRAE  thermal preference \cite{handbook2009american}                                                                        & Choose one (cooler, no change, warmer)                                                                                                                         \\ \hline
Clothing level                                                                       & Commonly used ASHRAE clothing insulation                   \cite{handbook2009american}                                                                                       & Choose multiple                                                                                                                                                \\ \hline
Seating location                                                                      & Seating location in the classroom                                                                                      & Click one point                                                                                                                                                \\ \hline
\begin{tabular}[c]{@{}l@{}} Behavioural/Emotional/ \\ Cognitive engagement\end{tabular} & \begin{tabular}[c]{@{}l@{}}Adapted In-class Student Engagement \\ Questionnaires (ISEQ) \cite{fuller2018development}\end{tabular}                   & \begin{tabular}[c]{@{}l@{}}-2: strongly disagree, -1: somewhat disagree, \\ 0: neither agree nor disagree, 1: somewhat agree,\\ 2: strongly agree\end{tabular} \\ \hline
Arousal/Valence                                                                       & \begin{tabular}[c]{@{}l@{}}Commonly used affective dimensions from \\ the Photographic Affect Meter (PAM) \cite{pollak2011pam}\end{tabular} & Choose one photo                                                                                                                                               \\ \hline
Confidence level                                                                   & Confidence level of the response                                                                                & \begin{tabular}[c]{@{}l@{}}1: not confident, 2: slightly confident, 3: moderately \\ confident, 4: very confident, 5: extremely confident\end{tabular}         \\ \hline
\end{tabular}
\caption{Collected annotations from the questionnaires.}
\label{tab:annotations}
\end{table}

\begin{figure}
	\centering
	\subfigure[The question of seating location ]{\includegraphics[width=0.151\textwidth]{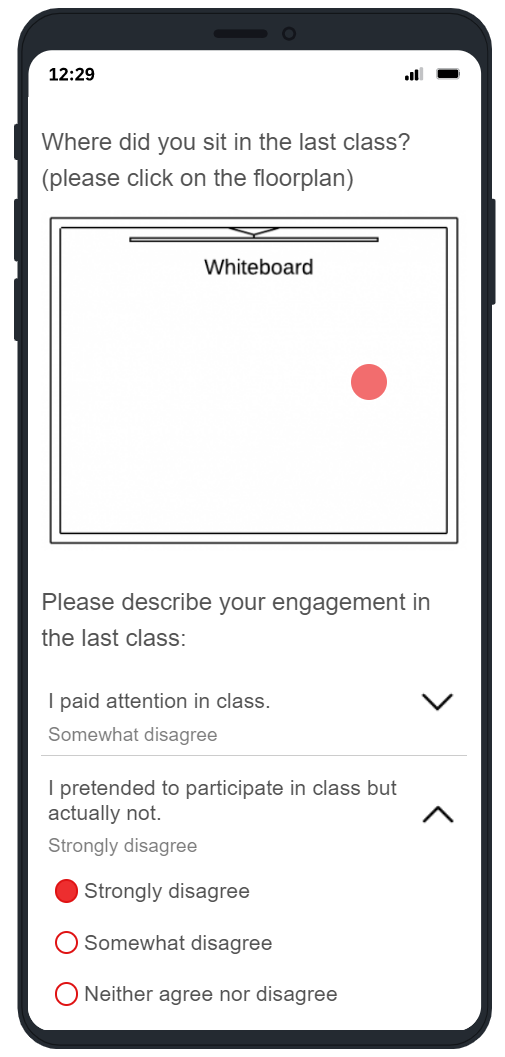}
	\label{fig:seatingscreen}}
	\hspace{2cm}
	\subfigure[PAM]{\includegraphics[width=0.15\textwidth]{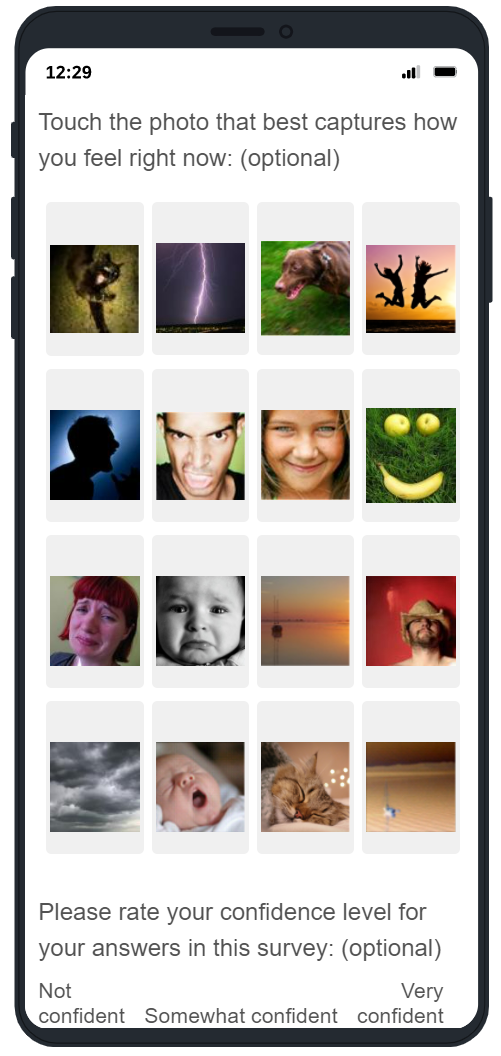}
    \label{fig:moodscrren}}
    \caption{Screenshots of the self-report survey.}
\end{figure}

\begin{figure}
	\centering
	\subfigure[Thermal Sensation \label{fig:heatmap}]{\includegraphics[width=0.31\textwidth]{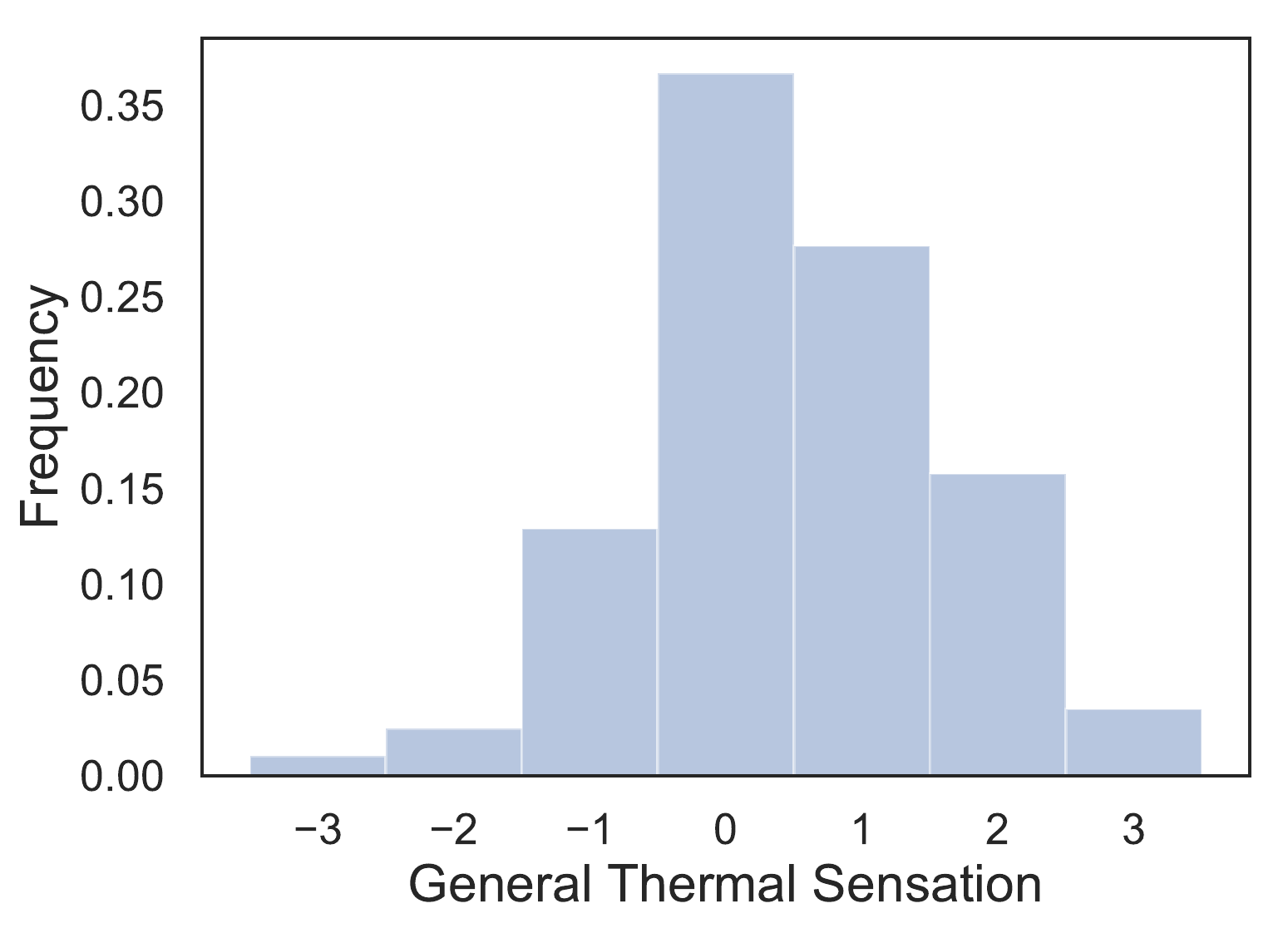}}
    \hspace{0cm}
	\subfigure[Thermal preference]{\includegraphics[width=0.24\textwidth]{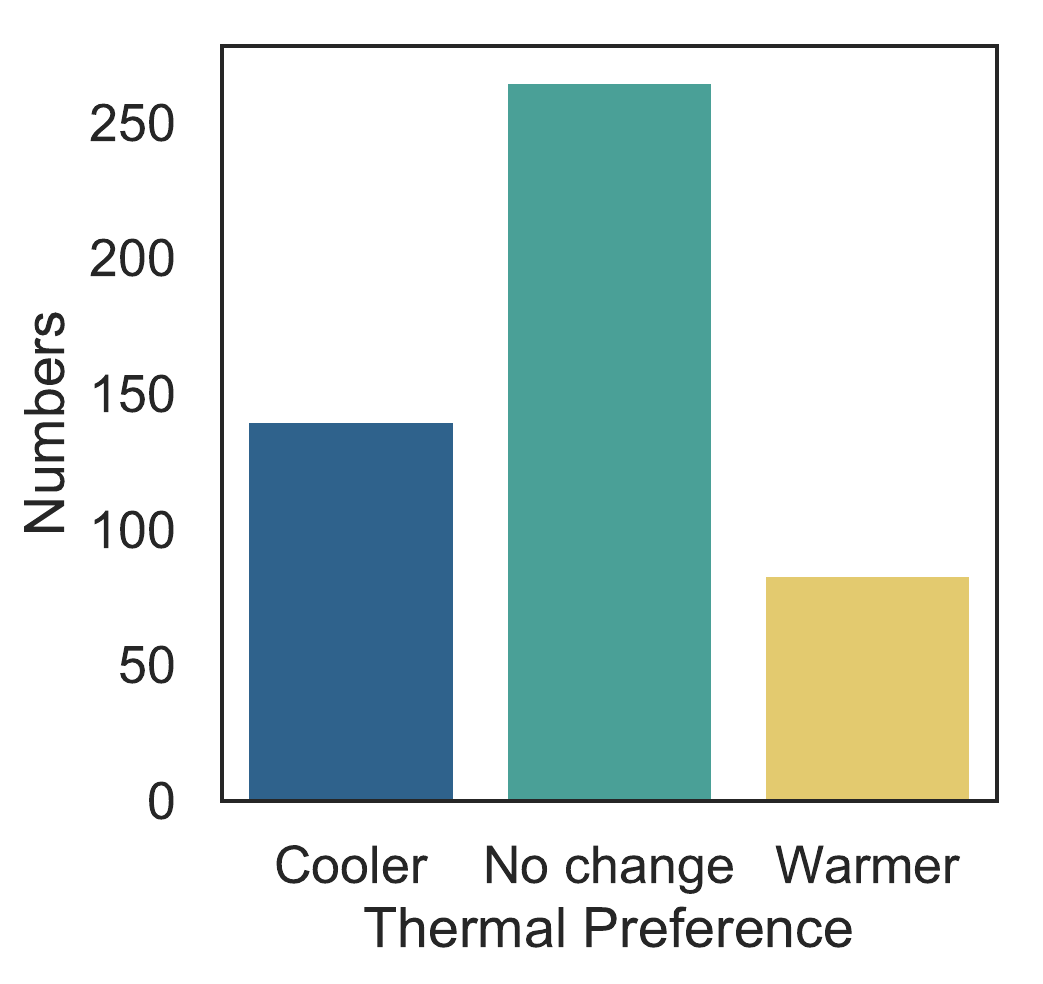}}
	\hspace{0cm}
	\subfigure[Clothing level\label{fig:gridmap}]{\includegraphics[width=0.34\textwidth]{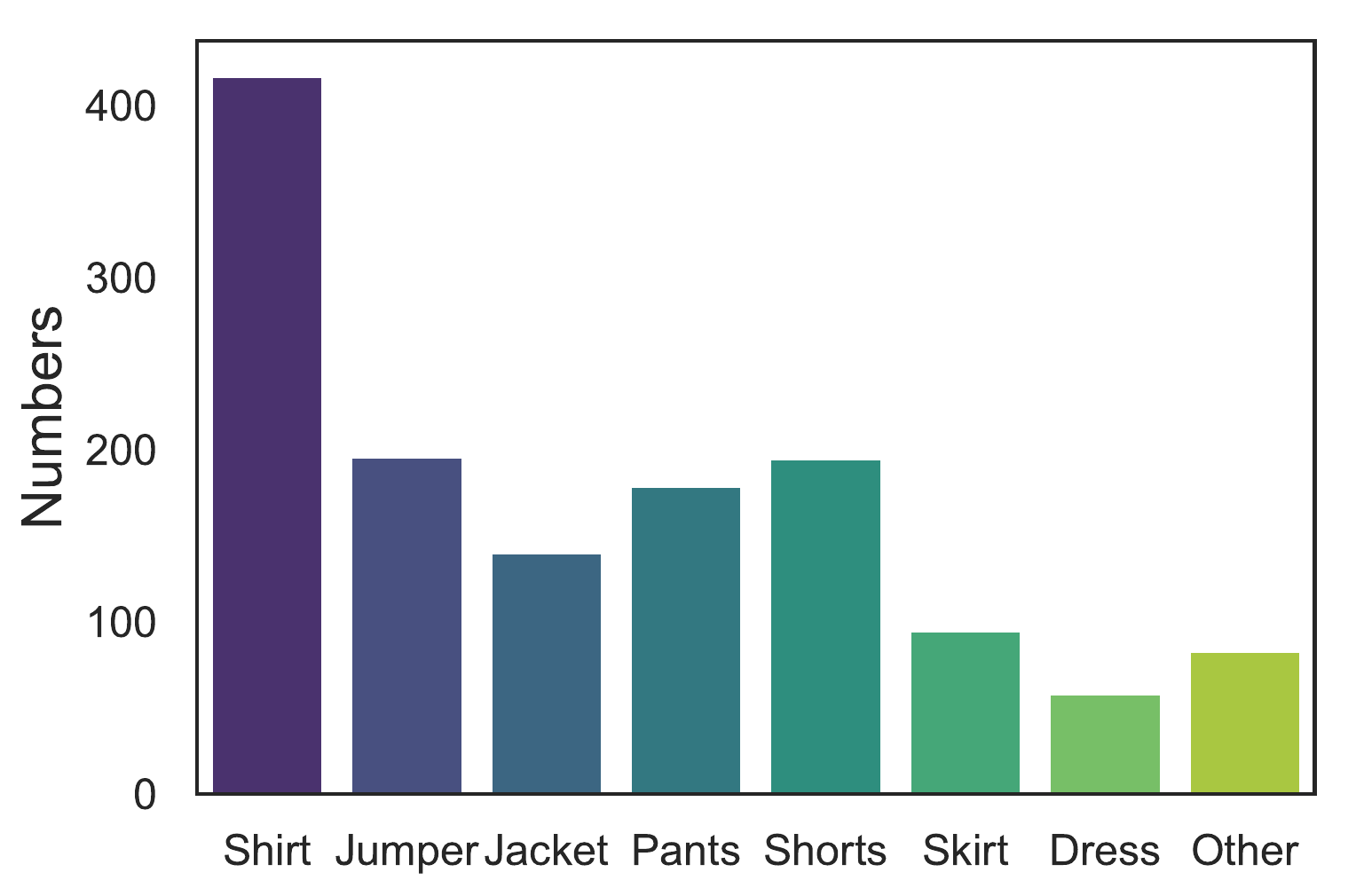}}
    \caption{Distribution of responses related to thermal comfort.}
    \label{fig: dis_sens}
\end{figure}

\begin{figure}
	\centering

	\subfigure[Multi-dimensional Engagement \label{fig:dis_emotion_eng:a}]{\includegraphics[width=0.28\textwidth]{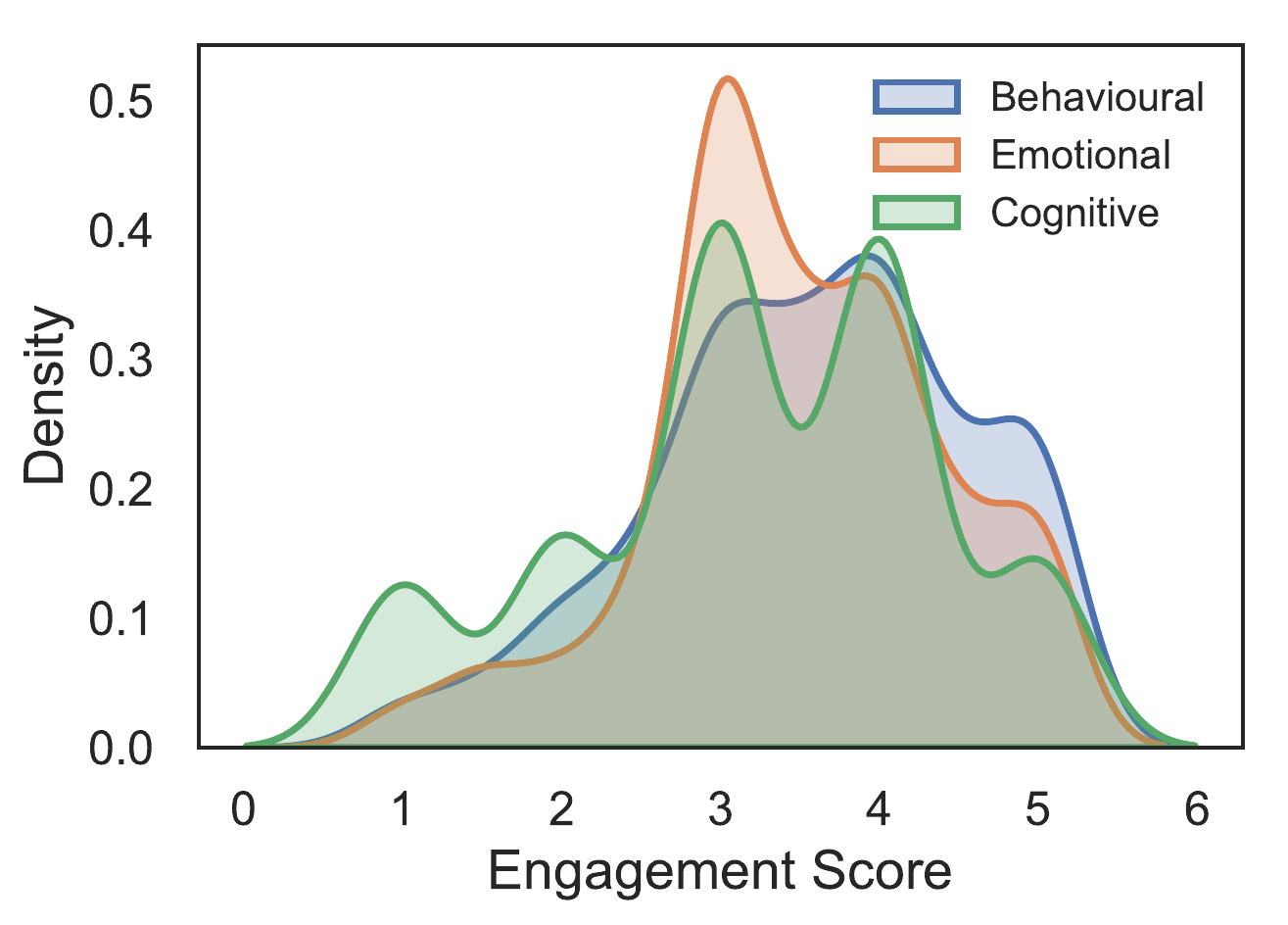}}
	\hspace{0cm}
    \subfigure[Overall Engagement across Participants \label{fig:dis_emotion_eng:c}]{\includegraphics[width=0.45\textwidth]{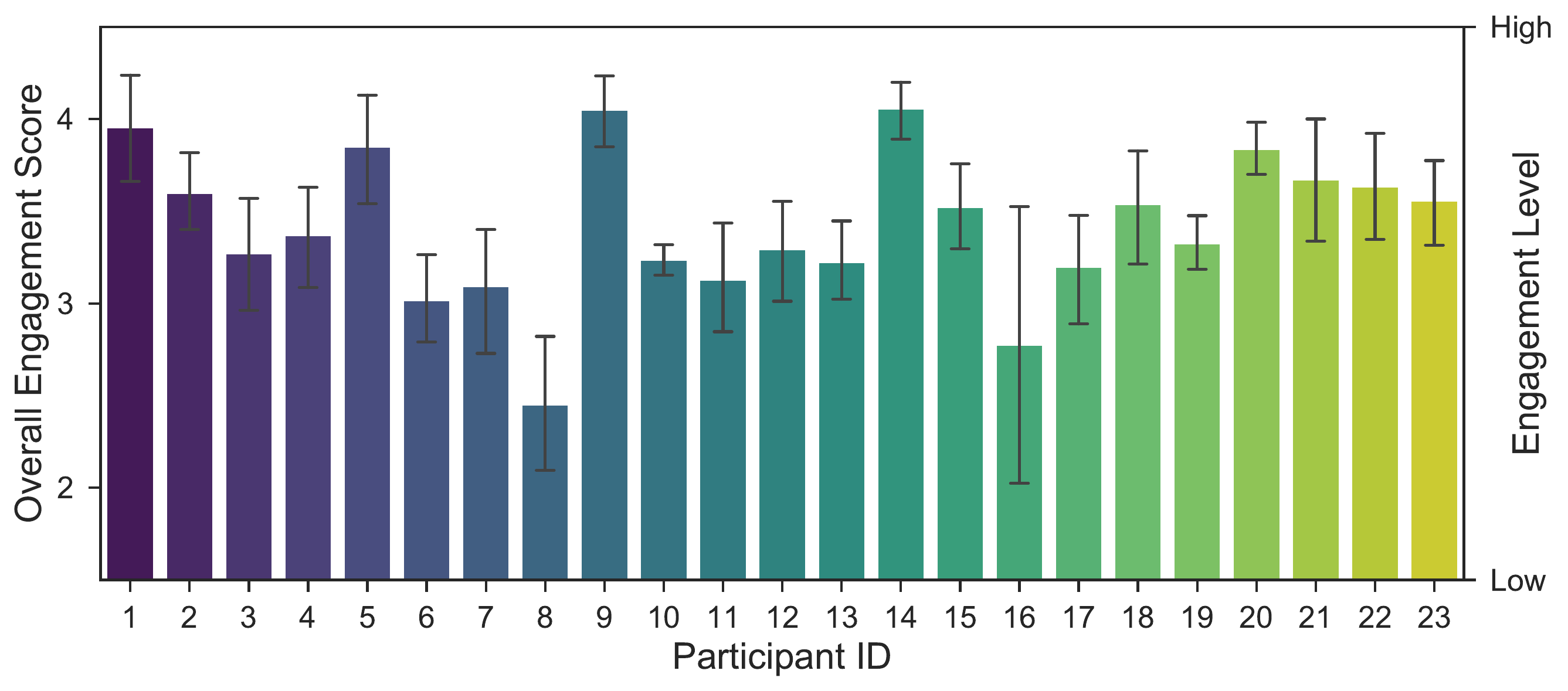}}
    \hspace{0cm}
\subfigure[Valence and Arousal \label{fig:dis_emotion_eng:b}]{\includegraphics[width=0.25\textwidth]{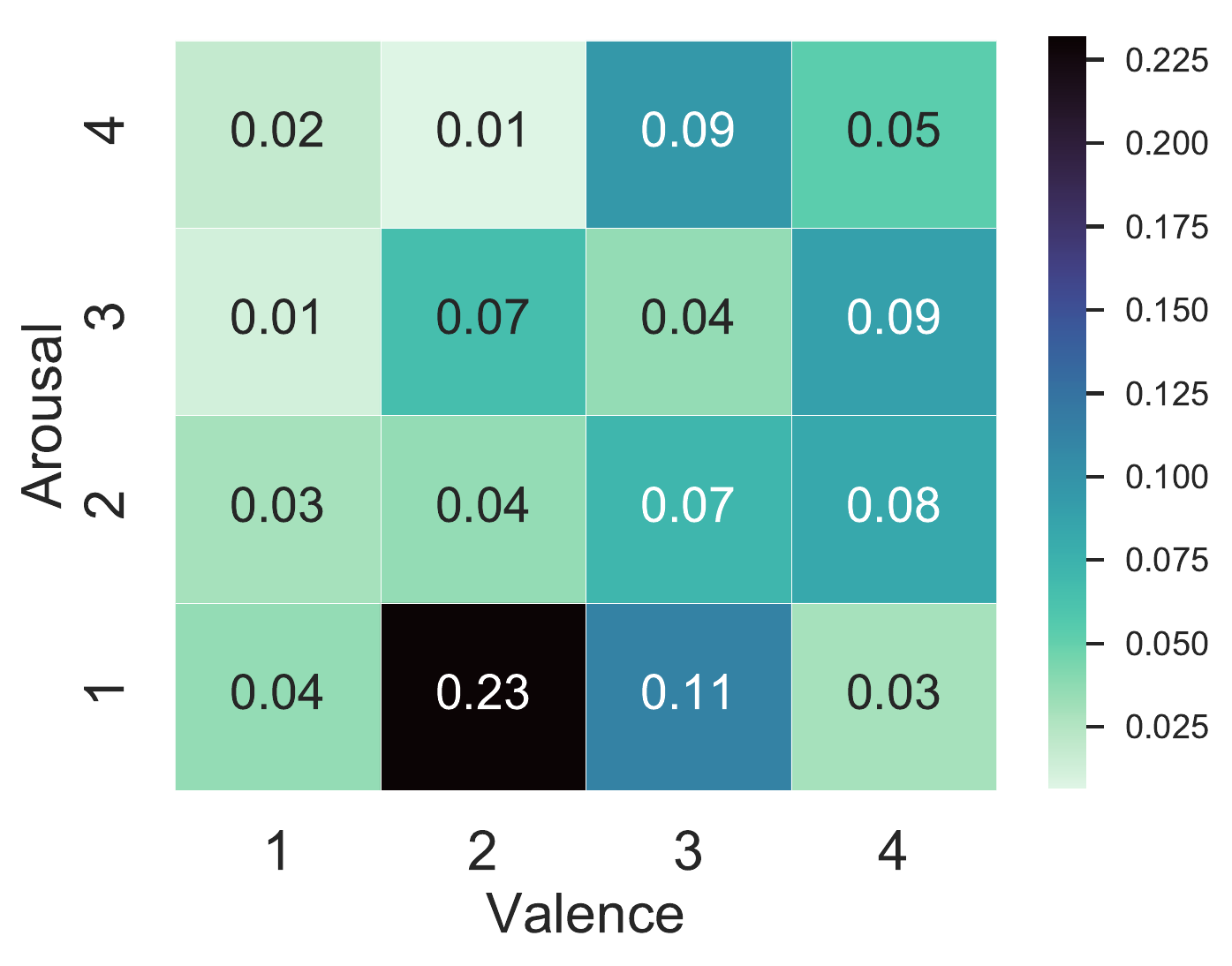}}
    \caption{Distribution of responses related to the engagement and emotion.}
    \label{fig: dis_emotion_eng}
\end{figure}

\begin{figure}
    \centering
    \includegraphics[width=0.9\textwidth]{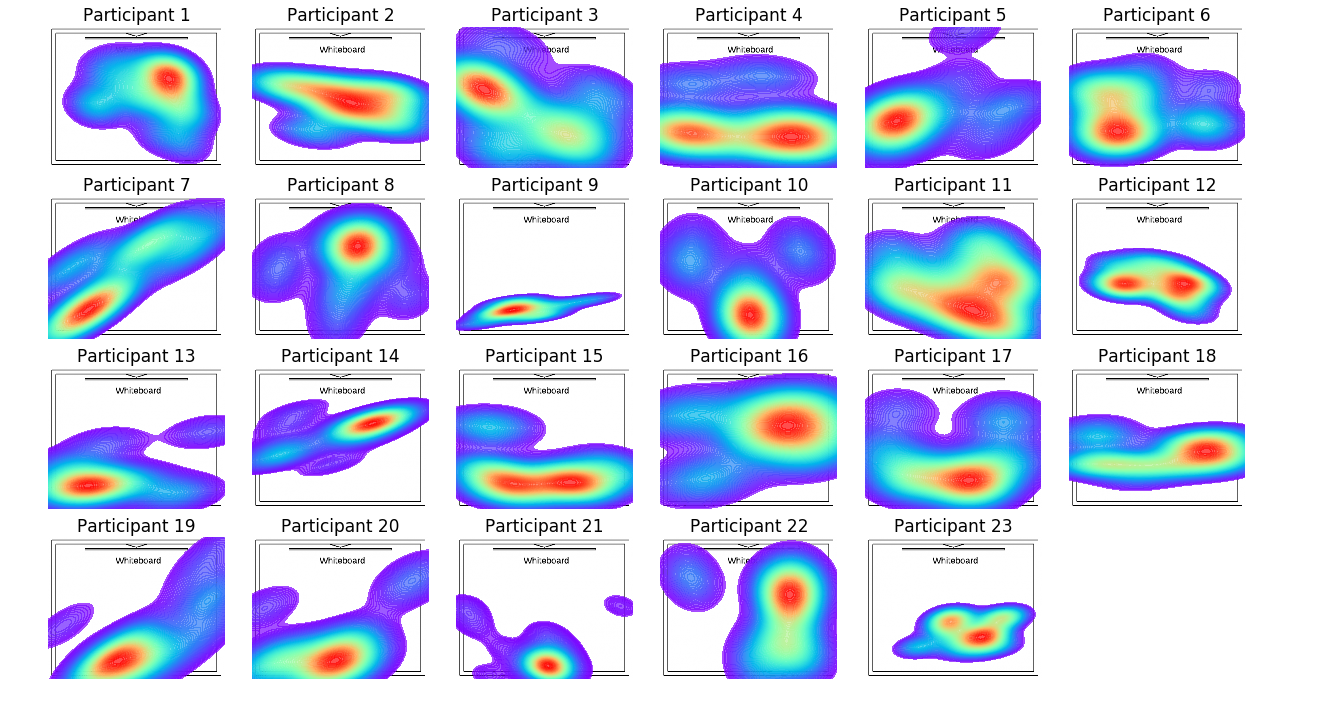}
    \caption{Distribution of seating locations across different participants.}
    \label{fig: locations}
\end{figure}

\textbf{Daily surveys.} On each school day, student participants were asked to complete online surveys (either through tablets placed in each classroom or using their own digital devices) at 11:00, 13:25 and 15:35 (directly after the second, fourth and fifth class). The length of the second and fourth class was either 40 min or 80 min, depending on the day of the week, and the fifth class always lasted 80 min. The curriculum in this school had a bi-weekly rhythm, i.e., the first and second weeks had different class schedules, but the first and third weeks were identical, as were the second and fourth weeks. The representative student was tasked with reminding the student participants to complete the online surveys on time, as described in Table \ref{tab:annotations}. The online questionnaire included 11 items related to the students' psychological states and behaviours (e.g., thermal comfort, student engagement and emotions). All the items (except the seating location and confidence level) were used directly or slightly adapted from the validated questionnaires widely used by researchers in this area. The screenshot of the question for seating location can be seen from Figure \ref{fig:seatingscreen}. Figure \ref{fig:moodscrren} illustrates the implemented PAM \cite{pollak2011pam} asking the user to choose one picture from a grid of 16 pictures in a library of 32 photos. Figure~\ref{fig: dis_sens} displays the distribution of responses to the thermal sensation (from -3 to 3), thermal preference and clothing level. The distribution of multidimensional (behavioural, emotional and cognitive) engagement can be found in Figure \ref{fig:dis_emotion_eng:a}, and the overall engagement across participants is reflected in Figure \ref{fig:dis_emotion_eng:c}. Figure \ref{fig:dis_emotion_eng:b} depicts the distribution of emotions in the valence and arousal dimensions. The numbers indicate the percentage frequencies, and the darker the colour, the higher the frequency of the specific emotion (e.g., arousal = 1 and valence = 2). Figure \ref{fig: locations} shows the distribution of seating locations across different participants.


\begin{figure}
	\centering
	\subfigure[Hour of the day \label{fig:self_report_dis1}]{\includegraphics[width=0.33\textwidth]{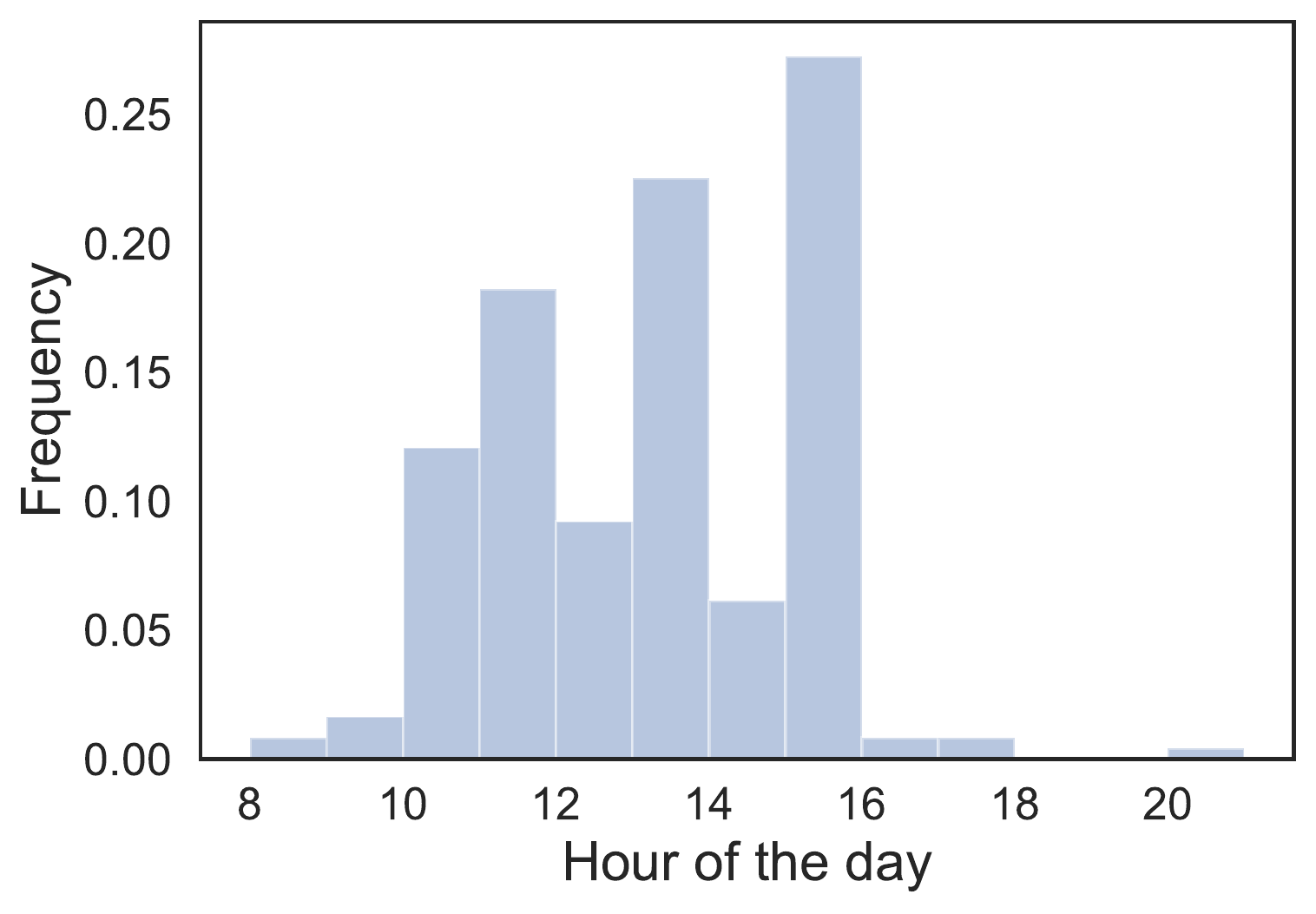}}
    \subfigure[Day of the week \label{fig:self_report_dis2}]{\includegraphics[width=0.33\textwidth]{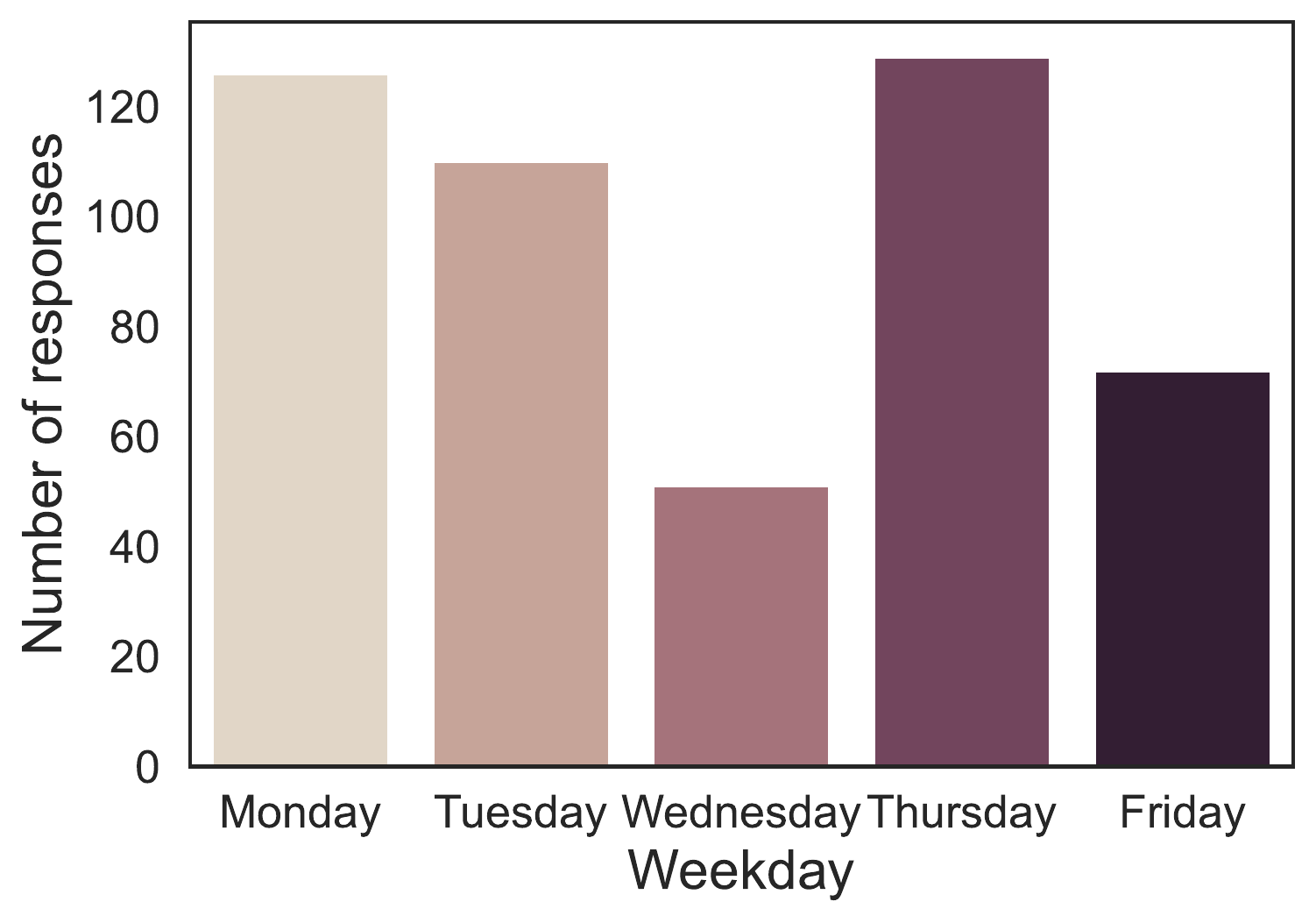}}
    \caption{Distribution of the survey responses over hours of the day or day of the week.}
    \label{fig: self_report_dis}
\end{figure}

Figure \ref{fig:self_report_dis1} shows the distribution of survey responses throughout the day. As students were requested to submit their self-reports directly after the second, fourth, and fifth classes (i.e., 11:00, 13:25 and 15:35, respectively), most responses were submitted at 11:00--12:00, 13:00--14:00 and 15:00--16:00. The survey responses that were recorded before the start of the targeted class or after the start of the next class should be removed if we aim to explore student engagement in the targeted class. However, if we wish to study thermal comfort, clothing insulation, emotion or confidence level, all the survey responses can be kept. Next, we explored the distribution of the survey responses through the week (see Figure \ref{fig:self_report_dis2}). We found that most students submitted their responses on Monday or Thursday. The number of students who submitted self-reported data on Wednesday was the lowest. The potential reason for this may be that students forgot to submit their responses, as they took breadth studies on Wednesday (normal studies on the other weekdays).

\begin{figure}
	\centering
	\subfigure[Heart Rate]{\includegraphics[width=0.43\textwidth]{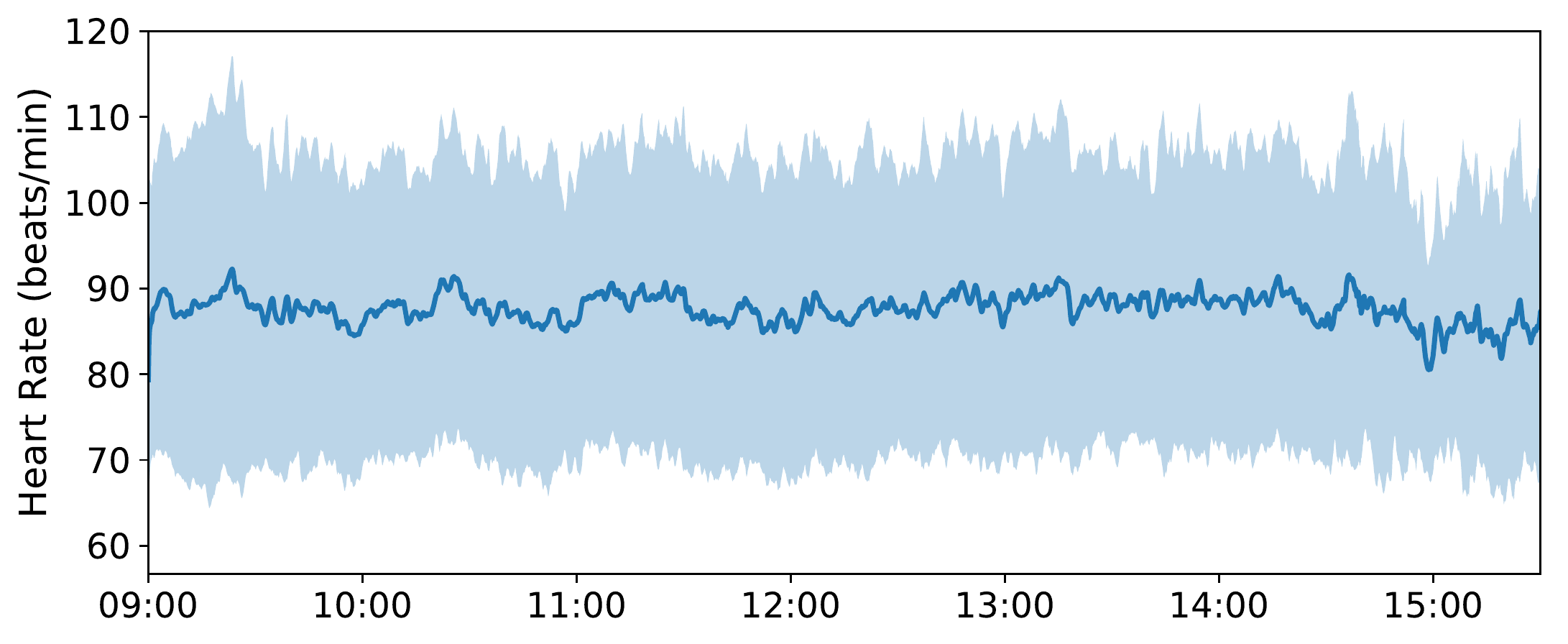}
	\label{fig: e4_hr}}
	\hspace{0cm}
	\subfigure[Electrodermal Activity]{\includegraphics[width=0.43\textwidth]{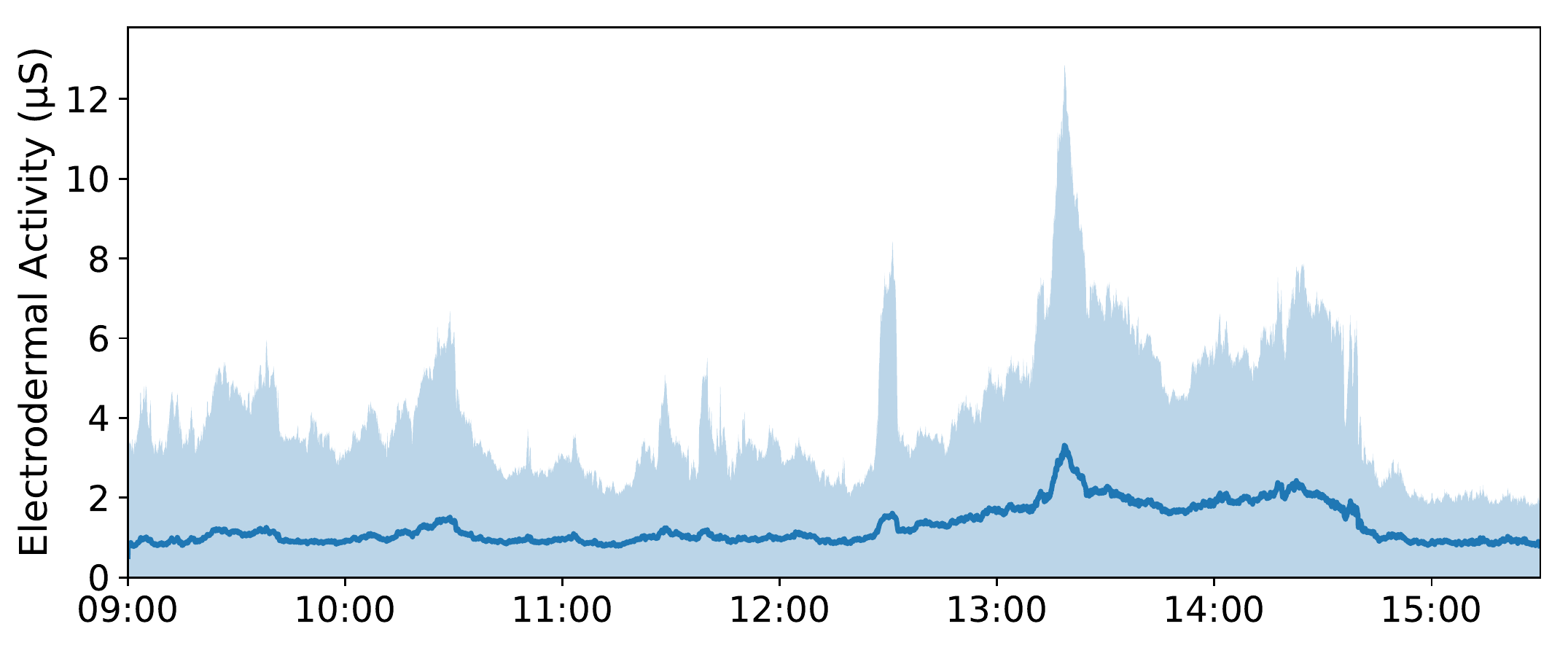}
    \label{fig: e4_EDA}}
    \subfigure[Skin Temperature]{\includegraphics[width=0.43\textwidth]{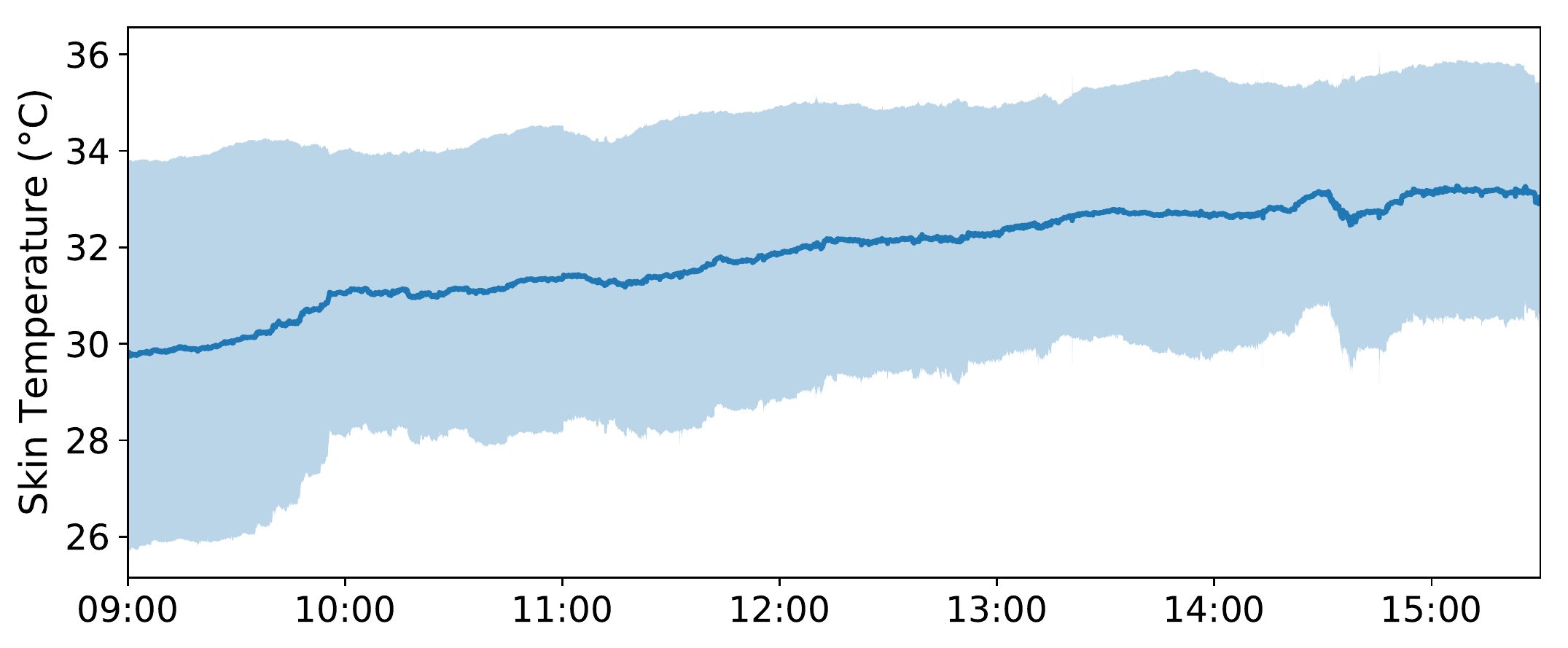}
    \label{fig: e4_ST}}
    \subfigure[3-axis Acceleration (Magnitude)]{\includegraphics[width=0.43\textwidth]{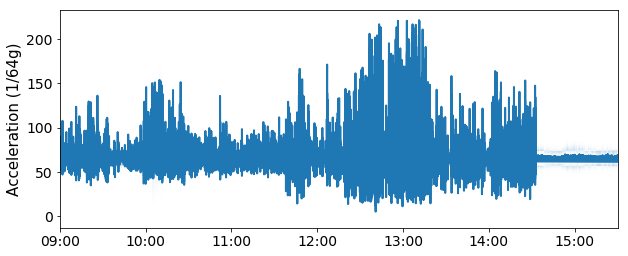}
    \label{fig: e4_ACC}}
    \hspace{0.2cm}
    \caption{Wearable signals per school day for all participants (369 traces in total).}
    \label{fig:e4_dis}
\end{figure}

\textbf{Empatica E4 wristband.} These wristband sensors (see Figure~\ref{fig: sub e4band}) were first proposed for use in studies by Garbarino et al. \cite{garbarino2014empatica}. These watch-like devices have multiple sensors: an EDA sensor, a photoplethysmography (PPG) sensor, a three-axis accelerometer (ACC) and an optical thermometer. EDA refers to constantly fluctuating changes in the electrical properties of the skin at 4 Hz; when the level of sweat increases, the conductivity of the skin increases. PPG sensors measure the blood volume pulse (BVP) at 64 Hz, from which the interbeat interval (IBI) and heart rate variability (HRV) can be derived. The ACC records in the range of [-2 g, 2 g] at 32 Hz and captures motion-based activity, which has been widely used in smartphones, wearables and other IoT devices \cite{gao2019predicting}. The optical thermometer reads peripheral skin temperature (ST) at 4 Hz. In recording mode, E4 wristbands can store 60 hours of data in memory, with a battery life of over 32 hours. They are lightweight, comfortable and waterproof, and were thus especially suitable for the continuous and unobtrusive monitoring of the participants in our study. Before the data collection, all wristbands were synchronised with the E4 Manager App, using a single laptop to ensure that the internal clocks were accurate. Each student was assigned a wristband sensor marked with their unique study ID. The students were asked to wear the wristband on their non-dominant hand, and to avoid pressing the button or performing any unnecessary movements during class. The teacher participants were only required to wear the wristbands while teaching the year 10 classes. Figure~\ref{fig:e4_dis} displays the distribution of wearable signals per school day for all participants. The blue line indicates the average values of signals calculated from 369 traces during school time (9:00 to 15:30). 

\textbf{DigiTech XC0422.} We set up two outdoor weather stations on-site: one in the prevailing NNW windward direction located at some distance from the buildings, and one on the SSE leeward side. These logged the data types shown in Table \ref{tab:max1} at 5-minute intervals via the school's guest WiFi to WUnderground.com where it can be accessed remotely. Note that these weather stations log solar irradiance values in W/m2 but only have a luminosity sensor. The method of conversion from lux to W/m2 is unclear from the product’s datasheet, but we assumed that it was in line with a commonly used, simplified conversion rate (e.g. Michael, 2019) \cite{michael2020conversion}.

\textbf{Netatmo Healthy Home Coach.} We collected indoor environmental data using Netatmo Healthy Home Coaches installed in 17 classrooms as shown in Figure~\ref{fig: sub netamo} and Figure~\ref{fig: sub room}. The logging specifications of these devices are shown in Table~\ref{tab:max2}. These devices measure indoor temperature, relative humidity, CO$_2$ levels and noise levels at a five-minute logging frequency. The data was uploaded in real-time via the school's guest WiFi to the Netatmo cloud platform from which we accessed the data remotely through our Netatmo account login. The analysed classrooms differed from one another in several aspects including the room geometry and orientation, as well as the number and location of windows. The placement of environmental sensor devices was therefore determined on a case-by-case basis, with the goal of finding an optimal trade-off between several, partly conflicting considerations, most of which were suggested by Wagner et al. \cite{wagner2018exploring}. For example, we tried placing the sensors close to the occupants but at the same time avoiding the sensors from being obstructed, biased or obtrusive due to their proximity to the occupants, furniture, heating elements, vents or appliances. The ASHRAE Standard 55 recommends temperature sensor heights of 0.1, 0.6 and 1.1 m for ankles, waists and heads of seated occupants, respectively. Given these guidelines, since only one device was installed per room in this study and the head height of children is lower than that of adults, we attempted to place the sensors at approximately 0.9 m. Three of the classrooms with a Netatmo station were rooms frequented by Year 10 students, therefore this data could be used in combination with the data captured by the E4 wristband sensors in the cross-sectional study. It should be noted that we started using these three devices and the outdoor weather stations before beginning the full longitudinal field study in all classrooms.

\textbf{Philio Temperature/Humidity Sensor.} The classrooms had split-system remote-controlled air-conditioning units for heating and cooling. We inferred their usage by measuring temperature fluctuations with Philio Temperature/Humidity Sensors placed at the outlets of the vents of the remote-controlled room air conditioning units. The sensors logged data via Z-Wave to Vera Edge hubs, several of which were placed throughout the school due to their limited range. Data was logged at 5-minute intervals using custom LUA scripts via the VeraAlerts app within the Vera SmartHome app, which enabled sending the data to the Pushbullet online platform from where they could be accessed remotely.

\begin{table}[]
\centering
\small
\renewcommand{\arraystretch}{1.1}
\begin{tabular}{|l|l|l|l|l|}
\hline
Type                    & Units & Range    & Accuracy                                                                                      & Resolution                                                                                      \\ \hline
Dry bulb temperature  & °C             & -40°C--60°C    & ± 1\%                                                                                                   & 0.1°C                                                                                                    \\\hline
Dew point temperature & °C             & -40°C--60°C    & ± 1\%                                                                                                   & 0.1°C                                                                                                    \\\hline
Relative humidity    & \%             & 1\%--99\%      & ± 5\%                                                                                                   & 1\%                                                                                                       \\\hline
Wind speed            & m/s            & 0 m/s--50 m/s    & \begin{tabular}[c]{@{}l@{}}±1 m/s (< 5 m/s)\\     ± 10\% ($\geq$ 5 m/s)\end{tabular} & 0.1 m/s                                                                                                   \\\hline
Gust speed           & m/s            & 0 m/s--50 m/s    & \begin{tabular}[c]{@{}l@{}}±1 m/s (< 5 m/s)\\     ± 10\% ($\geq$ 5 m/s)\end{tabular} & 0.1 m/s                                                                                                   \\\hline
Wind direction        & °              & 0°--360°       & ± 22.5°                                                                                                 & 22.5°                                                                                                    \\\hline
Rainfall              & mm             & 0 mm--9,999 mm    & ± 10\%                                                                                                  & \begin{tabular}[c]{@{}l@{}}0.3 mm (< 1000 mm)\\   1 mm ($\geq$ 1000 mm)\end{tabular} \\\hline
Light                 & lux            & 0 lux--400,000 lux & ± 15\%                                                                                                  & 0.1 lux                                                                                                   \\\hline
Solar radiation      & W/m$^2$           & N/A                & N/A                                                                                                    & N/A \\ \hline
\end{tabular}
\caption{DigiTech XC0422 logging specifications.}
\label{tab:max1}
\end{table}
\begin{table}[]
\small
\centering
\renewcommand{\arraystretch}{1.1}
\begin{tabular}{|l|l|l|l|l|}
\hline
Type                     & Units & Range  & Accuracy                                                                                               & Resolution \\ \hline
Dry bulb temperature & °C             & 0°C--50°C   & ± 0.3°C                                                                                                        & 0.1°C              \\\hline
Relative humidity    & \%             & 0\%--100\%     & ± 3\%                                                                                                          & 1\%                \\\hline
CO$_2$                  & ppm            & 0 ppm--5,000 ppm  & \begin{tabular}[c]{@{}l@{}}±50 ppm (\textless  1,000 ppm)\\     ± 5\% ($\geq$ 1,000 ppm)\end{tabular} & 1 ppm               \\\hline
Noise             & dB             & 35 dB--120 dB & N/A                                                                                                               & 1 dB                \\ \hline
\end{tabular}
\caption{Netatmo Healthy Home Coach logging specifications.}
\label{tab:max2}
\end{table}

\subsection*{Data post-processing}
For the longitudinal study, we extracted the environmental data from their respective online platforms and rounded each data point’s timestamp to the nearest 5-minute step to enable the aggregation of data from different sources, and interpolated over missing data points. The data from the two weather stations were averaged for each time step. In cases where one of the stations had missing data, we used the other weather station's data point. The outdoor wind direction was originally given in a 16-step scale of cardinal directions which we converted to numerical angle values in degrees.

\begin{figure}
    \centering
    \includegraphics[width=0.7\textwidth]{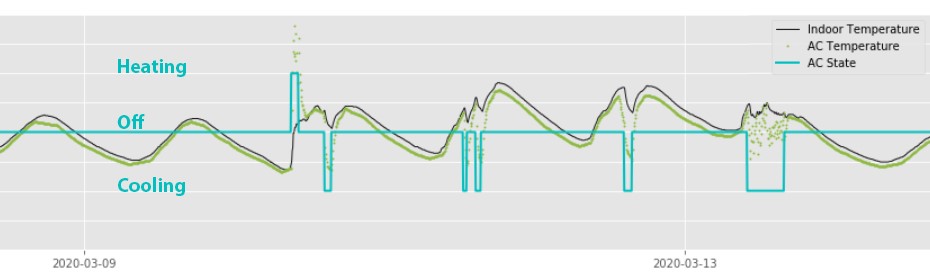}
    \caption{Data sample showing the indoor ambient temperature, the temperature reading at the air conditioning vent and the inferred air conditioning states.}
    \label{fig:max3}
\end{figure}

Within the context of this field study, there was no way to directly monitor when the air conditioning units were in use. Instead, we measured their use indirectly with the Philio temperature sensors mounted to the air conditioning outlets. Creating an algorithm that reliably distinguishes all four event types (cooling switched on, cooling switched off, heating switched on and heating switched off) is a task that would have exceeded the scope of our research. Instead, we used threshold values of the temperature slope to predict events. If the current state was off, then a sudden rise would be classified as switching on the heating; if the current state was cooling, the same rise in temperature would be classified as switching off the cooling. We found that classifying a temperature difference of 0.5 degrees from one 5-minute time step to the next to be useful in auto-detecting the majority of switching events. However, this crude method was limited in its predictive capability, not least because in an 'on' state, an air-conditioner will automatically keep switching itself off and on in order to remain within a certain temperature band around the selected set point; an example of this can be seen in the last air-conditioning period shown in Figure \ref{fig:max3}. Therefore, we relied on a visual assessment of the data and manually overwrote time frames with states that appeared to have been incorrectly categorised by the algorithm, based on the curves of the indoor temperature and AC temperature graphs. This is a potential source of error, but we assumed that the assessment was sufficiently accurate for this study - an assumption that proved correct when testing it on site. 
We aggregated all the data into spreadsheets for each classroom individually, and added several data, including columns that identified holiday periods, occupancy and time frames of insufficient data coverage.

\begin{figure}
    \centering
    \includegraphics[width=1\textwidth]{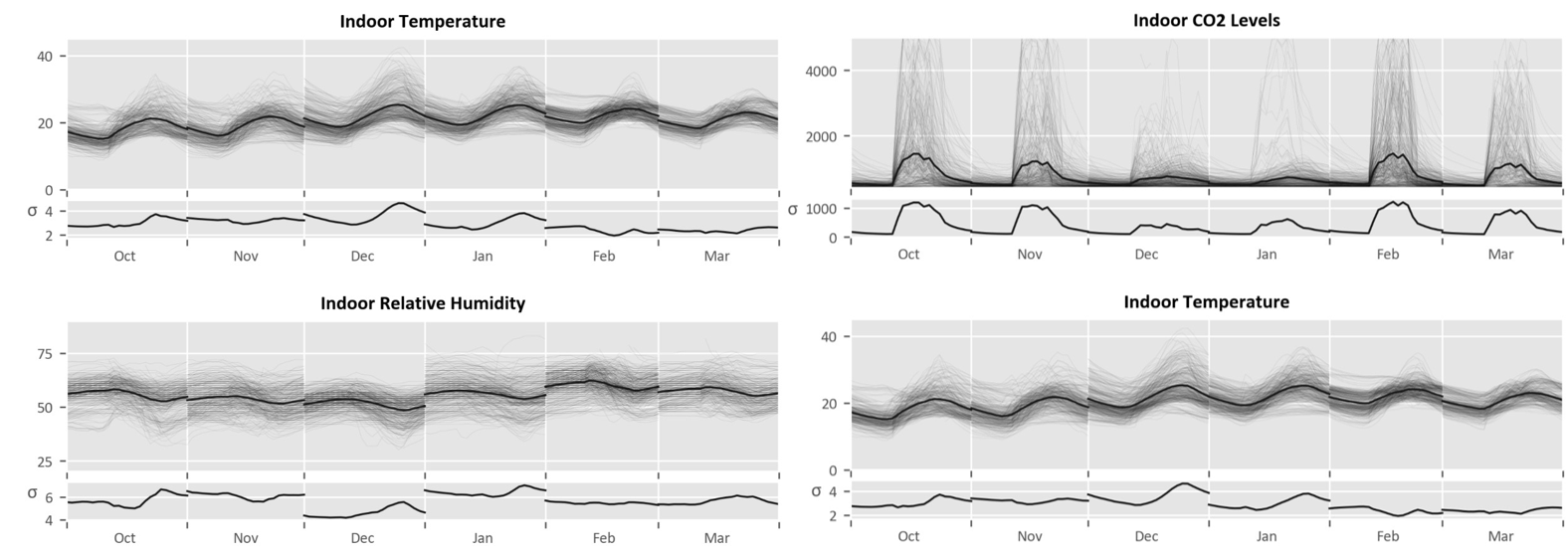}
    \caption{Daily indoor environmental trends by month.}
    \label{fig:max5}
\end{figure}

In the cross-sectional study, for the wearable data, we converted the timestamps of the wristband sensor readings from raw time intervals and Unix time to the local date-time format. Then we categorized the wearable data based on a different date. The wearable data were extracted according to the scheduled length for the second, fourth and fifth class which ends at 11:00, 13:25, 15:35. For the online survey, we received a total of 488 valid online surveys from students with a response rate of 35.3\%. We also received 22 online surveys from teachers. Then, we aligned the survey data to one of the three classes. We set the survey responses before 11:25 pm to belong to the second class, between 12:15 pm--14:15 pm to belong to the fourth class, after 14:15 pm to belong to the fifth class.

\section*{Data Records}



\subsection*{Summary}
The data are available on the \textit{Figshare} data repository \cite{gao_marschall_burry_watkins_salim_2021}. It includes two elements: \textit{In-Gauge} dataset for the longitudinal study and \textit{En-Gage} dataset for the cross-sectional study.  For the \textit{En-Gage} dataset, we have provided two versions: the original raw data by date and organised data based on the different class groups of the participants.

The \textit{In-Gauge} dataset consists of comma-separated variable (CSV) files - one for each classroom. Each classroom's spreadsheet contains time-related information and outdoor weather conditions (these are obviously identical for all classrooms). Furthermore, each classroom has information on its own indoor climate, whether or not it is occupied according to the class schedule, and information on whether its room air-conditioner is in heating or cooling mode. The \textit{En-Gage} dataset includes physiological signals measured with the wristband sensors as well as self-reported engagement, thermal comfort, seating locations, and emotion data from the student and teacher participants. 

\begin{figure}
    \centering
    \includegraphics[width=0.9\textwidth]{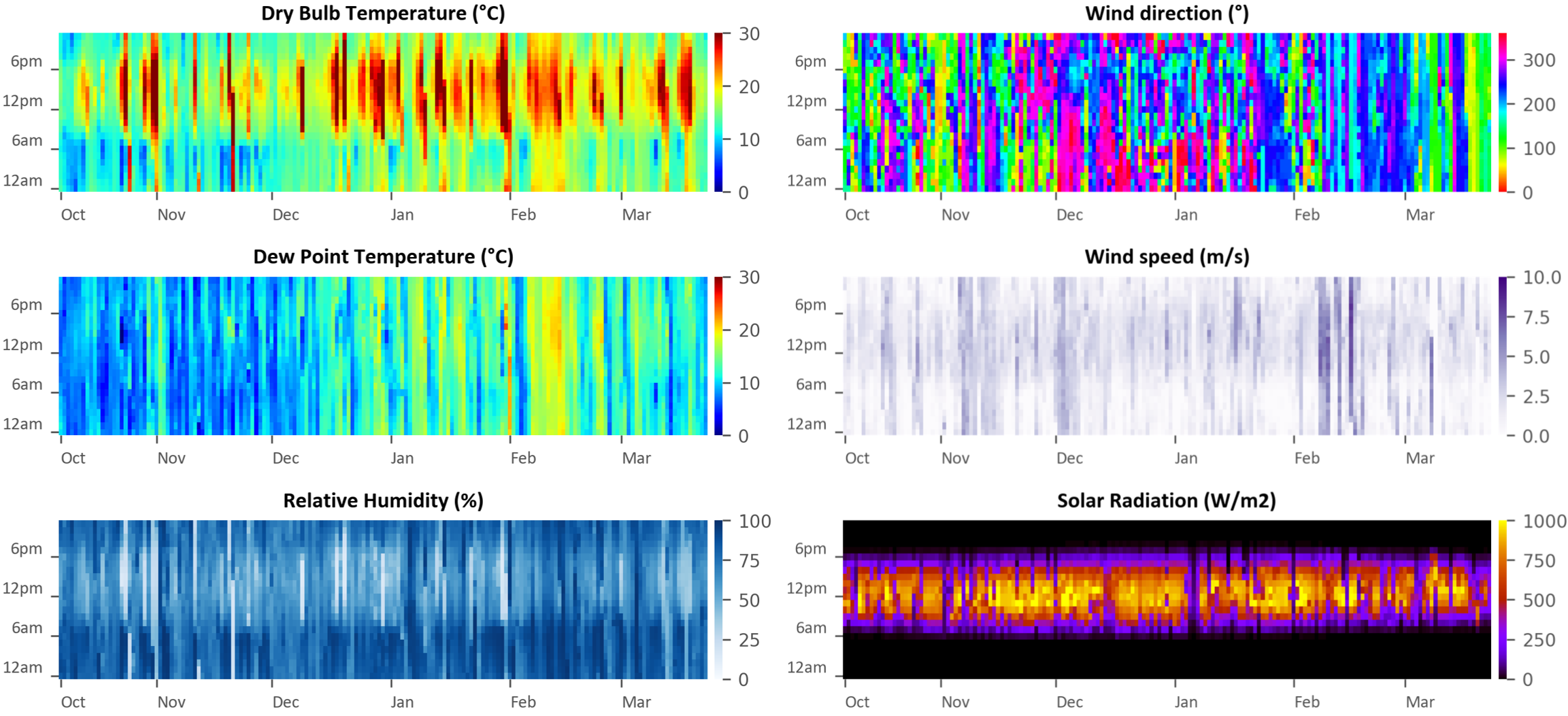}
    \caption{Hourly outdoor climate (averaged between the two weather stations).}
    \label{fig:max4}
\end{figure}

\subsection*{Contents}
In the following, we describe the directories and files in our datasets.

\subsubsection*{Longitudinal}
This folder contains all data pertaining to the longitudinal field study. It consists of a TXT file describing the dataset and 16 CSV files (one for each classroom). The CSV file names correspond to the classroom names. Each CSV file has a single header line and each of the following rows contains the following timestamped data at a resolution of 5 minutes per row:

\begin{itemize}[topsep=0pt,itemsep=-0.9ex]
    \item Timestamp: Local datetime format e.g. '2019-10-08 18:25:00'.
    \item Year:	An integer of either 2019 or 2020.
    \item Month: An integer between 1 and 12.
    \item DayOfYear: An integer between 1 and 365.
    \item Occupied:	'0' means that the room was not occupied at this time according to the classroom schedule; '1' means it was.
    \item SchoolDay: '0' means that this day was not a school day; '1' means it was.
    \item Hour: An integer representing the hour od day from 0 to 23.
    \item LessonNumber: An integer signifying which class is currently taking place (note that each school day started with a 10-minute assembly referred to here as the '0'th class): '-1' = outside of school hours; '0' = 8:50--9:00; '1' = 9:00--9:40; '2' = 9:40--10:20; '3' = 10:20--11:00; '4' = 11:25--12:05; '5' = 12:05--12:45; '6' = 12:45--13:25; '7' = 14:15--14:55; '8' = 14:55--15:35; '9' = Recess times or special "Breadth Studies" session on Wednesdays.
    \item LessonPct: A fraction between 0.0 and 1.0 describing how much of the current class has passed.
    \item IndoorTemperature: A decimal number representing the current indoor temperature in °C.
    \item IndoorHumidity: An integer representing the current indoor relative humidity in \%.
    \item IndoorCO2: An integer representing the current indoor CO$_2$ concentration in ppm.
    \item IndoorNoise: An integer representing the current indoor noise level in dB.
    \item OutdoorTemperature: A decimal number representing the current outdoor temperature in °C.
    \item OutdoorHumidity: An integer representing the current outdoor relative humidity in \%.
    \item OutdoorDewpoint: A decimal number representing the current outdoor dewpoint temperature °C.
    \item OutdoorWindDirection: An integer representing the current outdoor wind direction in degrees, from 0 to 360 (0° = north wind, 90° = east wind, etc.).
    \item OutdoorWindSpeed: A decimal number representing the current outdoor wind speed in m/s.
    \item OutdoorGustSpeed: A decimal number representing the current outdoor gust speed in m/s.
    \item Precipitation: A decimal number representing the current outdoor precipitation in mm.
    \item UvLevel: An integer between 0 and 11 representing the current outdoor Global Solar UV Index.
    \item  SolarRadiation: An integer representing the current outdoor solar radiation intensity in W/m$^2$.
    \item CoolingState: '0' means that the room air-conditioner was currently not cooling the room; '1' means it was.
    \item HeatingState: '0' means that the room air-conditioner was currently not heating the room; '1' means it was.
    \item UsabilityMask: For timeframes where too much data was missing, we set this UsabilityMask field to "False" for the entire day. During holidays, the UsabilityMask also reads "False".
\end{itemize}

\begin{figure}
    \centering
    \includegraphics[width=1\textwidth]{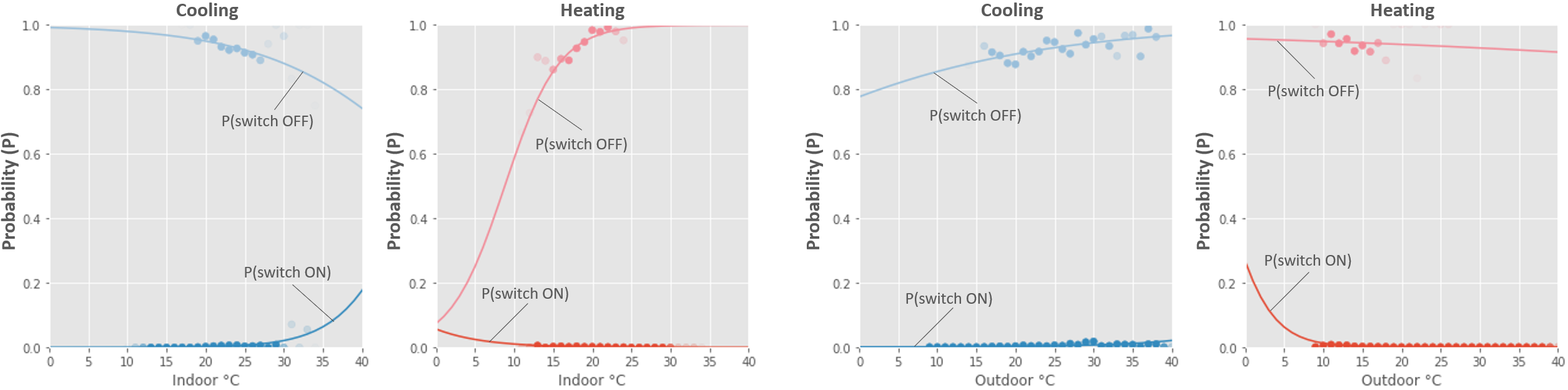}
    \caption{An example for an application of the dataset: creating models of occupant behaviour to predict the switching of air-conditioners, for both heating and cooling. The above are examples using simple logistic regression; the left ones use indoor temperature as the independent variable, the right ones use outdoor temperature as the independent variable.}
    \label{fig:max6}
\end{figure}

\subsubsection*{Participant\_class\_info}
This folder contains demographic information on the background questionnaires participants, and the class schedule. Note that for several survey questions, we adopted the 5-point Likert scale: -2 = 'strongly disagree', -1 = 'somewhat disagree', 0 = 'neither agree nor disagree', 1 = 'somewhat agree' and 2 = 'strongly agree'. The \textit{Participant\_class\_info} folder contains the following files:

\begin{enumerate}[topsep=1pt,itemsep=-0.5ex]
    \item \textit{Student.csv}. Each row in this file contains a participant ID (\textit{Column A}), gender (\textit{Column B}), age in years (\textit{Column C}), form room, math room and language room (\textit{Columns D--F}), and three background questions (\textit{Columns G--K}) related to their general thermal comfort and engagement in class. Specifically, \textit{Columns G to I} represent, respectively, the questions 'What is your general feeling in the classroom?' [-3 = cold, -2 = cool, -1 = slightly cool, 0 = neutral, 1 = slightly warm, 2 = warm, 3 = hot], 'When I am engaged in class, I usually don't feel too hot or too cold' and 'When I am engaged in class, I could get distracted when the room is too hot or too cold'. For the latter 2 questions, we adopted the 5-point Likert scale.
    
    \item \textit{Teacher.csv}. Each row in this file contains a participant ID (\textit{Column A}), gender (\textit{Column B}), age in years (\textit{Column C}), teaching subject (\textit{Columns D}), and three background questions similar to the \textit{student.csv} file, except that we changed the last two questions slightly from 'When I am engaged in class, [...]' to 'When I am engaged in teaching, [...]'.
    
    \item \textit{Class\_table.csv}. We generate this file from the class schedule obtained from the school. Each row in this file contains the information of one single class, including the unique class ID (\textit{Column A}), classroom (\textit{Column B}), date (\textit{Column C}), start time of the current class (\textit{Column D}), finish time of the current class (\textit{Column E}), length of the class (\textit{Column F}), week (\textit{Column G}), weekday (\textit{Column H}), the order of the class (\textit{Column I}) and the course name (\textit{Column J}). Specifically, \textit{Column K} shows whether students take this class in a form group, where '0' indicates they are not in a form group, 'all' indicates all students take this class in one whole form group (i.e., Assembly, Chapel), the R1/R2/R3 means students take this class in form groups and their form room is R1, R2 or R3.
\end{enumerate}

\begin{figure}
    \centering
    \includegraphics[width=0.64\textwidth]{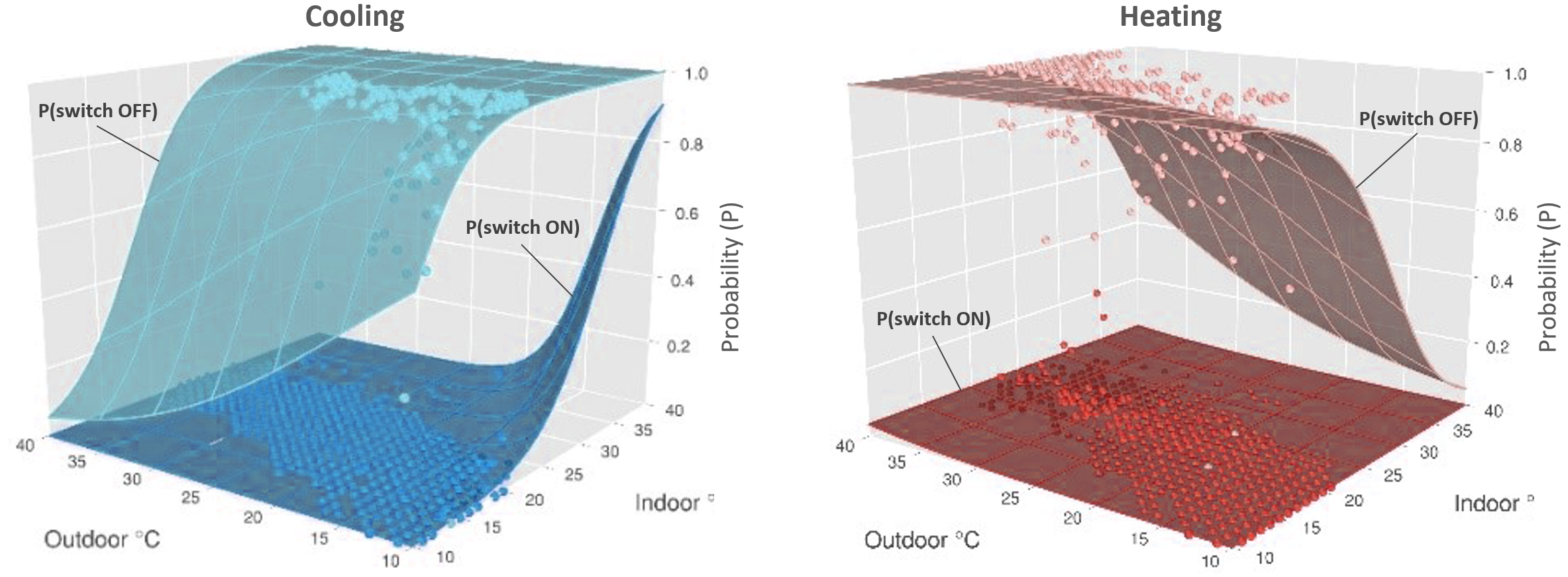}
    \caption{An example for an application of the dataset: creating a model of occupant behaviour to predict the switching of air-conditioners, for both heating and cooling. This example model uses multiple logistic regression, with indoor and outdoor temperatures as the independent variables. }
    \label{fig:max7}
\end{figure}

\subsubsection*{Survey}
This folder contains 2 files: \textit{Student\_survey.csv} and \textit{Teacher\_survey.csv}. 

\textit{Student\_survey.csv} contains the 488 survey responses including 15 columns where \textit{Column A} is participant ID and \textit{Column B} is the recorded time. There are columns containing thermal comfort-related information (\textit{Columns C--G}), multi-dimensional student engagement (\textit{Columns H--L}), mood (\textit{Column M}), and confidence level of the survey (\textit{Column N}). The engagement questions were rated using the Likert-scale. To calculate the engagement score, users should reverse the responses in item 2 and item 4, then calculate the average of the 5-point Likert scale for each dimension of engagement. The specific columns relate to the following questions:

\begin{itemize}[topsep=0pt,itemsep=-0.9ex]
    \item \textit{Column C: Thermal\_sensation}: "How do you feel right now in the classroom?" [-3 = cold, -2 = cool, -1 = slightly cool, 0 = neutral, 1 = slightly warm, 2 = warm, 3 = hot].
    
    \item \textit{Column D: Thermal\_preference}: "Would you like to be?" [Cooler, No change, Warmer].

    \item \textit{Column E: Clothing}: "What are you wearing now? (multiple options allowed)" [Shirt, Jumper, Jacket, Pants, Shorts, Skirt, Dress, Other].
    
    \item \textit{Columns F--G: Loc\_x, Loc\_y}: "Where did you sit in the last class? (please click on the floorplan)" [x, y pixels in the  400*321 room thumbnail where x = y = 0 at the upper left corner].

   \item  \textit{Columns H--L: Engage\_1, 2, 3, 4, 5}: "Please describe your engagement in the last class": [I paid attention in class], [I pretended to participate in class but actually not], [I enjoyed learning new things in class], [I felt discouraged when we worked on something], [I asked myself questions to make sure I understood the class content].

\item \textit{Columns M--N: Arousal, Valence}: "Touch the photo that best captures how you feel right now (optional)" [We assigned the arousal and valence values from 1--4 to each picture. For instance, for the right bottom picture, valence = 4 and arousal = 1].

\item \textit{Column N: Confidence\_level}: "Please rate your confidence level for your answers in this survey (optional)" [5-point Likert scales where 1 = Not confident, 2 = Slightly confident, 3 = Moderately confident, 4 = Very confident, 5 = Extremely confident].
\end{itemize}

\textit{Teacher\_survey.csv} contains the 22 survey responses by the teachers. The file includes 11 columns where \textit{Column A} is the recorded time, \textit{Column B} is the wristband ID, \textit{Columns C--E} are the thermal comfort-related information, \textit{Columns F--G} are the engagement related information, and \textit{Column K} is the confidence level of the survey. For the wristband ID in \textit{Column B}, A/B/C/D represent the classrooms R1/R2/R3/R4. The specific columns relate to the following questions:

\begin{itemize}[topsep=0pt,itemsep=-0.9ex]
\item \textit{Column B: Wristband\_id}: "Please enter your wristband ID." [A, B, C, D].

\item \textit{Column C: Thermal\_sensation}: "How do you feel right now in the classroom?"  [-3 = cold, -2 = cool, -1 = slightly cool, 0 = neutral, 1 = slightly warm, 2 = warm, 3 = hot].

\item \textit{Column D: Thermal\_preference}: "Would you like to be?" [Cooler, No change, Warmer].

\item  \textit{Column E: Clothing}: "What are you wearing now? (multiple options allowed)" [Shirt, Jumper, Jacket, Pants, Shorts, Skirt, Dress, Other].

\item \textit{Columns F--G: Engage\_1, 2, 3, 4, 5}: "Please describe your engagement in the last class": [I was excited about teaching], [I felt happy while teaching], [While teaching, I paid a lot of attention to my work], [I cared about the problems of my students], [I was aware of my students' feelings].

\item \textit{Column K: Confidence\_level}: "Please rate your confidence level for your answers in this survey (optional)" [5-point Likert scales where 1 = Not confident, 3 = Somewhat confident, 5 = Very confident].

\end{itemize}

\subsubsection*{Raw\_wearable\_data}
This folder includes 20 sub-folders named with the date of data collection (e.g., '20191122'), containing the raw wearable data for each day during the 4-week data collection. In each sub-folder, there are multiple sessions from different participants. Some participants provided more than 1 session on the same day. The name of each session consists of two parts connected by an underscore: the unique session ID and the participant ID. For example, the session named '1567380164\_18' indicates the data is provided by participant 18. There are 6 \textit{CSV} files in each session, and each of these files (except \textit{IBI.csv}) has the following format: the first row is the initial time of the session expressed as a Unix timestamp in UTC. The second row is the sample rate expressed in Hz. Specifically:

\begin{enumerate}[itemsep=-0.5ex]

\item \textit{ACC.csv} contains data from a 3-axis accelerometer sampled at 32 Hz which is configured to measure accelerations in the range of [-2g, 2g]. Acceleration is the rate of change of the velocity with respect to time, where SI (International System of Units) \cite{international2001international} derived unit for acceleration is the metre per second squared ($m \cdot s^{-2}$) where 1g is equal to 9.80665 $m \cdot s^{-2}$. The unit in this file is 1/64 g where the raw value of 64 indicates 1g. The 3 columns refer to the x, y, and z-axis, respectively. 
   
\item \textit{BVP.csv} contains BVP signals sampled at 64 Hz which is the primary output from the PPG sensor. BVP signals can be used to compute the inter-beat-intervals (IBI) and heart rate (HR).

\item \textit{EDA.csv} contains data from an electrodermal activity (EDA) sensor expressed as micro siemens ($\mu{S}$) sampled at 4 Hz. The variation of EDA values indicates the electrical changes of the skin surface and the EDA arises when the skin receives nerve signals from the brain and sweat level increases \cite{braithwaite2013guide}.

\item \textit{HR.csv} contains the average heart rate data extracted from the BVP signals, calculated in spans of 10 seconds. The first row is the initial time of the session and it is 10 seconds after the beginning of the recording. The sampling rate of heart rate is 1 Hz.

\item \textit{IBI.csv} contains the time intervals between a participant's heartbeats extracted from the BVP signals. This file does not have a sampling rate. The first column is the time (with respect to the starting time) of the detected inter-beat interval expressed in seconds (s). The second column is the duration in seconds (s) of the detected inter-beat interval (i.e., the distance in seconds from the previous beat).

\item \textit{TEMP.csv} contains data from a temperature sensor expressed in degrees Celsius ($^{\circ}C$), sampled at 4 Hz.

\end{enumerate}

\begin{figure}
	\centering
	\subfigure[Raw signals \label{fig:quality:before}]{\includegraphics[width=0.46\textwidth]{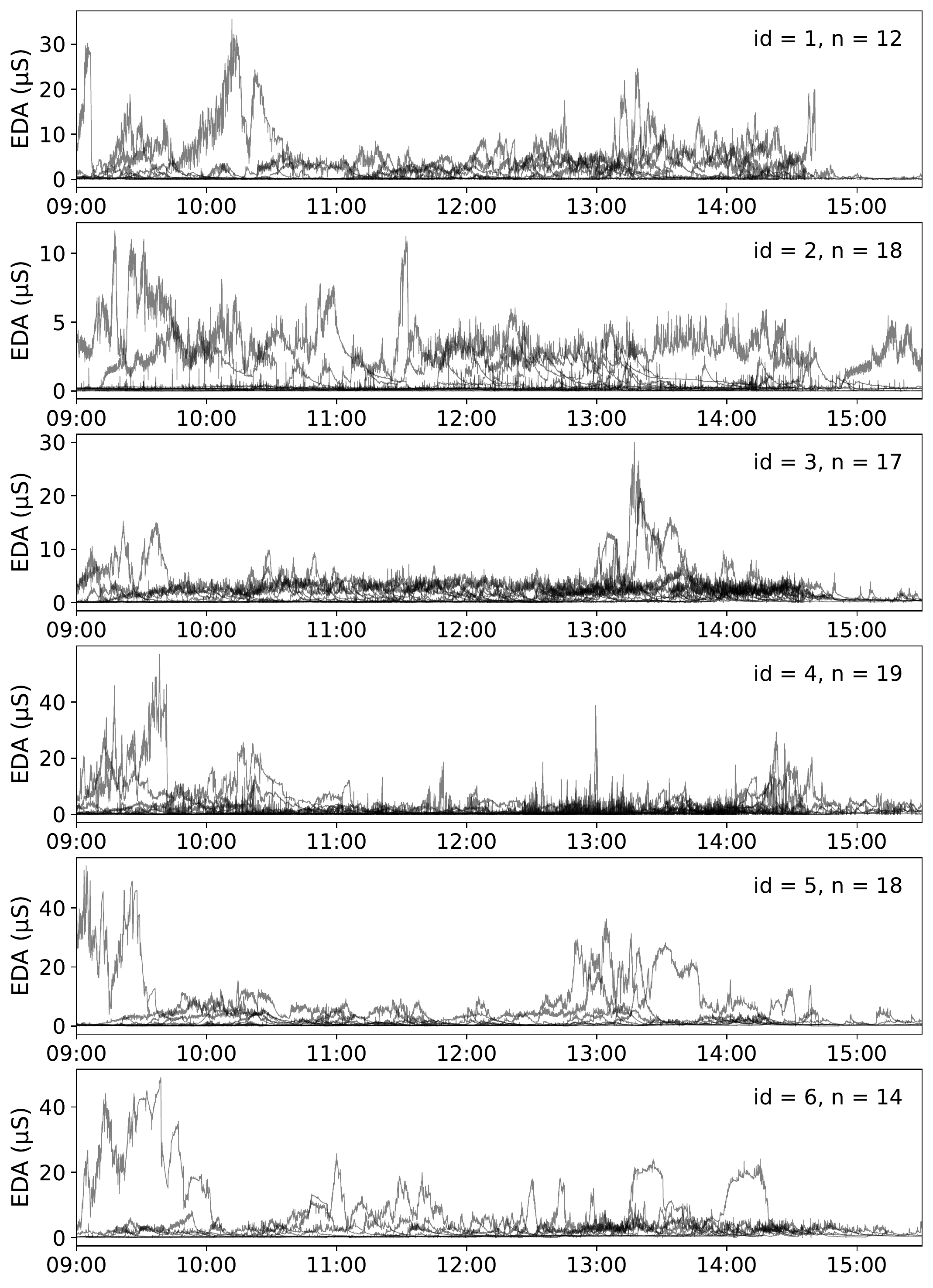}}
    \subfigure[Valid signals \label{fig:quality:after}]{\includegraphics[width=0.46\textwidth]{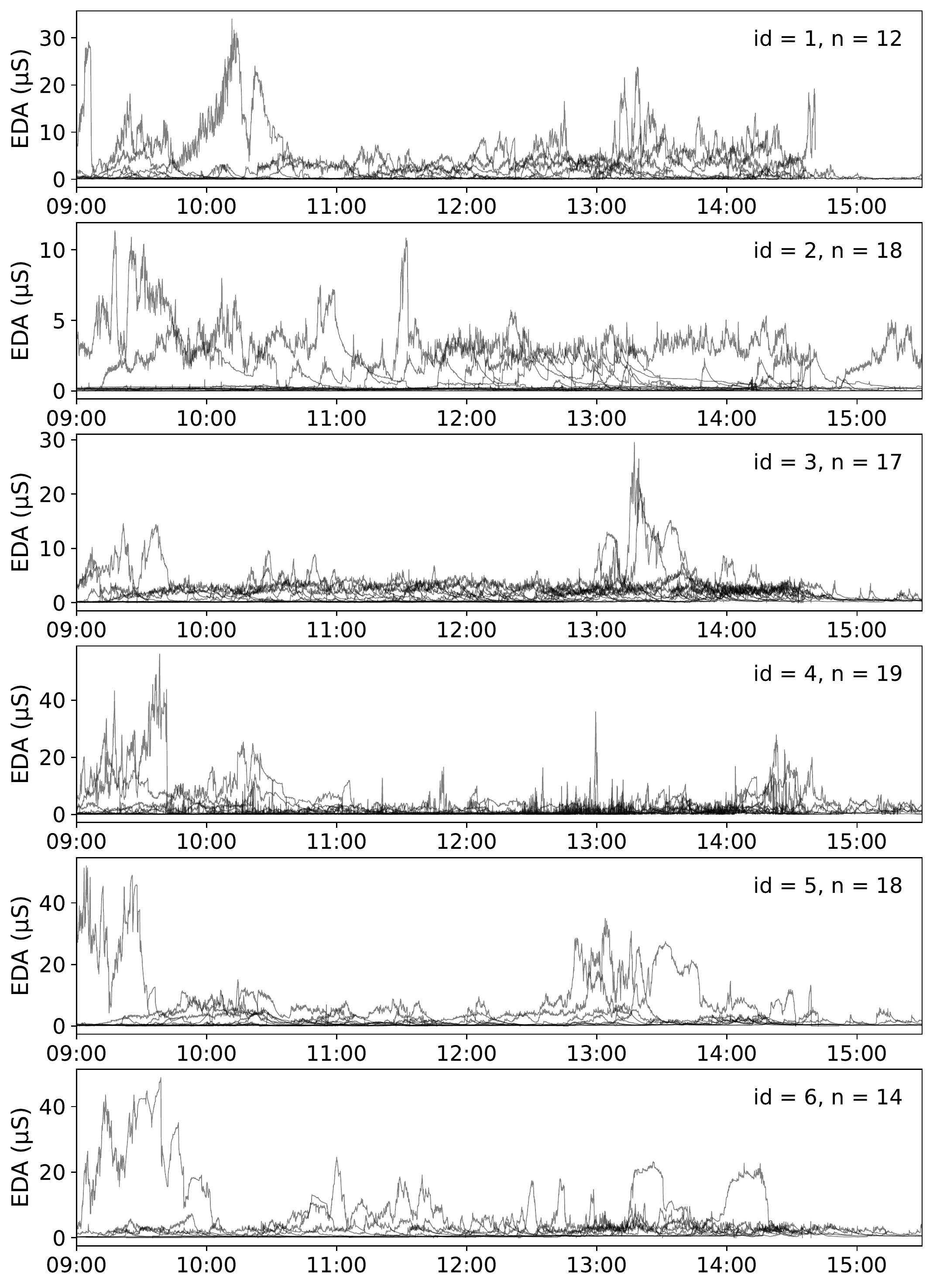}}
    \caption{EDA signals per school day for six participants before and after quality control.}
    \label{fig: quality}
\end{figure}

\subsubsection*{Class\_wearable\_data} 

The \textit{Class\_wearable\_data} folder contains 221 sub-folders representing 221 different classes during which the wearable data were recorded. Each sub-folder is named by the unique ’Class\_id’ as shown in the  \textit{Class\_table.csv}.
Each sub-folder includes further sub-folders named by the unique participant id or simply the label 'teacher’. These contain data from the wristband sensors for each participant of this class. There are 6 \textit{CSV} files in each sub-folder: \textit{ACC.csv}, \textit{EDA.csv}, \textit{BVP.csv}, \textit{HR.csv}, \textit{IBI.csv}, and \textit{TEMP.csv}. The format of these files is identical to the ones in the \textit{Raw\_wearable\_data} folder.

\section*{Technical Validation}
\textbf{Indoor and outdoor environmental data}.
Table \ref{tab:max1} and Table \ref{tab:max2} show the specifications of the outdoor and indoor weather stations, respectively. Figure \ref{fig:max3} shows a sample of data captured from these sensors and the air-conditioning states that we inferred from the data. Figure \ref{fig:max5} shows data captured by the indoor weather station. Figure \ref{fig:max4} shows data captured by the outdoor weather station. We were not able to find the accuracy of the Philio sensors in their datasheet, but assumed that it was sufficient for the purposes of our study. Figure \ref{fig:max6} and Figure \ref{fig:max7} demonstrate a potential use case for the longitudinal dataset; here, we fitted different logistic regression models to the air-conditioning usage data. Within the scope of our research, we were not able to validate these models. The figures serve merely as an illustration of what the dataset may be used for.

\textbf{Wearable signals}. In the dataset, wearable signals include EDA, HRV, skin temperature and 3-axis acceleration. The above signals can be utilized to understand people's behaviours, engagement, thermal comfort and emotion. Before the data analysis and modelling, the data cleaning stage needs to be conducted to remove noises and motion artefacts for quality control \cite{babaei2021critique}. Usually, there are several types of noises during the data collection from E4 wristbands \cite{gao2020n}: flat responses, abrupt signal drops and quantization errors. Flat responses (i.e., 0 micro siemens) happen when there is poor contact between skin and the wristband, and abrupt signal drops occur with movements of wristbands. The quality of wearable measurements in the dataset has been thoroughly examined.

\textit{Missing values}. Though physiological signals are mostly error-free for the majority of files in the dataset, a portion of data is missing due to issues of devices or human errors (e.g., poor contact, exhausted battery, unintentionally turning-off wristbands, unwilling to keep wearing wristbands). On average, participants contribute wearable signals for 15 days, with different wearing times during the school day (9:00--15:30). P13 and P16 contribute the least number of wearing days (\textit{n} = 6) and P21 wear the wristbands for most days (\textit{n} = 28). Then we calculate the missing values (i.e., 0 micro siemens for EDA signal) due to flat responses or motion artefacts. On average, 2.66\% of data is missing for all participants, with P10 having the maximal portion of missing values (27.37\%) and P20 having the least missing values (0.01\%).

\textit{Quality Control}. Figure \ref{fig: quality} shows the EDA signals in the school day for six participants before and after applying quality control. Due to space limitations, we only show the data from six participants where \textit{n} indicates the number of wearing days of wristbands. The quality control involved the removal of motion artefacts (MAs). The median filter with a 5-second window is applied on EDA signals per school day as in previous research \cite{gao2020n}. After quality control, the portion of missing values decreases from 2.66\% to 2.46\%. To better control the quality of wearable signals, various methods can be applied such as visual inspection \cite{gao2020n}, different filters with different settings, shape-based artefacts detection \cite{gashi2020detection}, etc. However, the aforementioned methods for different types of wearable signals (e.g., EDA, HRV) are usually selected or adapted according to the specific needs and purposes of researchers. Most importantly, unlike environmental data, there is no absolute ground truth for the physiological signals (EDA, HRV, and skin temperature) and physical activities (3-axis acceleration) of people in the real world. It does not help much to present a fixed and complete pipeline for the quality control of wearable signals. Therefore we only consider dealing with the missing values with a median filter for the EDA data validation as an example of quality control.

\section*{Usage Notes}

Our datasets include the outdoor/indoor/wearable sensing data and the self-report occupants' thermal comfort, learning engagement, and emotions while at school. This dataset is the first publicly available dataset for studying the daily behaviours and engagement of high school students using heterogeneous sensing. For the longitudinal outdoor and indoor sensing data, the most straightforward potential usage is to derive predictive models of how occupants operate room air-conditioning units \cite{schwee2019room}. Our dataset could potentially be useful to examine the relationships between indoor/outdoor climates and physiological signals of occupants, which provide opportunities for the future design of intelligent feedback systems to benefit both students and staff on campus.

Specifically, various data mining (e.g., segmentation \cite{deldari2020espresso}, clustering \cite{shao2016clustering,shao2019onlineairtrajclus}) and modelling techniques \cite{salim2020modelling,carlucci2020modeling,kjaergaard2020current} could be explored to build prediction models for measuring occupants' mental state using sensor-based physiological and behavioural recordings in buildings. This could be further used for various applications in future studies: (1) \textit{Monitoring signs of disengagement and negative emotions of students} \cite{gao2020n,rahaman2020ambient}. Measuring the study engagement and emotions of students is beneficial to both teachers and students. Teachers will be able to improve their teaching strategies to create the right learning environment, improve the learning experience for students and re-engage students with low engagement \cite{gashi2018using, gao2020n}. Students will be able to self-track their learning engagement and emotions, which could promote their self-regulation and reflective learning. (2) \textit{Studying peer effects in educational settings} \cite{sacerdote2014experimental}. It could be helpful to explore group-wise seating behaviours and their relationship to perceived engagement and physiological synchrony \cite{gao2021investigating}. (3) \textit{Providing comfortable indoor environments for occupants}. It is possible to mitigate the negative effects of hot weather on student learning by using air conditioning \cite{park2020heat}, and teachers could ventilate classrooms timely to prevent excess carbon dioxide from affecting students' concentration \cite{arief2017hoc,arief2018rup,arief2018scalable}.

Some limitations of the datasets need to be addressed. Firstly, Empatica E4 wristbands are susceptible to noises caused by many factors, e.g., loose electrodes and faulty wiring \cite{mccarthy2016validation}, and other devices such as indoor and outdoor weather stations may also be subject to similar systematic errors. Additionally, the accuracy of measuring physiological and behavioural data using E4 wristbands is limited. Menghini et al.  \cite{menghini2019stressing} found that similar accuracy could not be achieved when comparing EDA signals between wrists and fingers. Lead wire extension is one promising solution to improve the accuracy of E4 wristbands, which allows EDA recordings to be moved from the wrist to the surface of the fingers or palm, thus eliminating the potential site differences. In our data collection, using the E4 wristband is the best option for collecting data from student and teacher participants on-site without putting an extra burden on them. Finally, besides the quality control for EDA measurements, additional measures could be further employed to validate the collected data (e.g., self-report data and other physiological signals) through appropriately designed studies.

\section*{Code Availability}

Python code for prepossessing the data and implementing the segmentation based on different classes are available online \url{https://github.com/cruiseresearchgroup/InGauge-and-EnGage-Datasets}. 


\begin{thebibliography}{10}
\urlstyle{rm}
\expandafter\ifx\csname url\endcsname\relax
  \def\url#1{\texttt{#1}}\fi
\expandafter\ifx\csname urlprefix\endcsname\relax\def\urlprefix{URL }\fi
\expandafter\ifx\csname doiprefix\endcsname\relax\def\doiprefix{DOI: }\fi
\providecommand{\bibinfo}[2]{#2}
\providecommand{\eprint}[2][]{\url{#2}}

\bibitem{haldi2011impact}
\bibinfo{author}{Haldi, F.} \& \bibinfo{author}{Robinson, D.}
\newblock \bibinfo{journal}{\bibinfo{title}{The impact of occupants' behaviour
  on building energy demand}}.
\newblock {\emph{\JournalTitle{Journal of Building Performance Simulation}}}
  \textbf{\bibinfo{volume}{4}}, \bibinfo{pages}{323--338}
  (\bibinfo{year}{2011}).

\bibitem{rijal2011algorithm}
\bibinfo{author}{Rijal, H.~B.}, \bibinfo{author}{Tuohy, P.},
  \bibinfo{author}{Humphreys, M.~A.}, \bibinfo{author}{Nicol, J.~F.} \&
  \bibinfo{author}{Samuel, A.}
\newblock \bibinfo{title}{An algorithm to represent occupant use of windows and
  fans including situation-specific motivations and constraints}.
\newblock In \emph{\bibinfo{booktitle}{Building Simulation}},
  vol.~\bibinfo{volume}{4}, \bibinfo{pages}{117--134}
  (\bibinfo{organization}{Springer}, \bibinfo{year}{2011}).

\bibitem{schiavon2013dynamic}
\bibinfo{author}{Schiavon, S.} \& \bibinfo{author}{Lee, K.~H.}
\newblock \bibinfo{journal}{\bibinfo{title}{Dynamic predictive clothing
  insulation models based on outdoor air and indoor operative temperatures}}.
\newblock {\emph{\JournalTitle{Building and Environment}}}
  \textbf{\bibinfo{volume}{59}}, \bibinfo{pages}{250--260}
  (\bibinfo{year}{2013}).

\bibitem{schweiker2012verification}
\bibinfo{author}{Schweiker, M.}, \bibinfo{author}{Haldi, F.},
  \bibinfo{author}{Shukuya, M.} \& \bibinfo{author}{Robinson, D.}
\newblock \bibinfo{journal}{\bibinfo{title}{Verification of stochastic models
  of window opening behaviour for residential buildings}}.
\newblock {\emph{\JournalTitle{Journal of Building Performance Simulation}}}
  \textbf{\bibinfo{volume}{5}}, \bibinfo{pages}{55--74} (\bibinfo{year}{2012}).

\bibitem{langevin2015tracking}
\bibinfo{author}{Langevin, J.}, \bibinfo{author}{Gurian, P.~L.} \&
  \bibinfo{author}{Wen, J.}
\newblock \bibinfo{journal}{\bibinfo{title}{Tracking the human-building
  interaction: A longitudinal field study of occupant behavior in
  air-conditioned offices}}.
\newblock {\emph{\JournalTitle{Journal of Environmental Psychology}}}
  \textbf{\bibinfo{volume}{42}}, \bibinfo{pages}{94--115}
  (\bibinfo{year}{2015}).

\bibitem{reinhart2003monitoring}
\bibinfo{author}{Reinhart, C.~F.} \& \bibinfo{author}{Voss, K.}
\newblock \bibinfo{journal}{\bibinfo{title}{Monitoring manual control of
  electric lighting and blinds}}.
\newblock {\emph{\JournalTitle{Lighting Research \& Technology}}}
  \textbf{\bibinfo{volume}{35}}, \bibinfo{pages}{243--258}
  (\bibinfo{year}{2003}).

\bibitem{cheung2019analysis}
\bibinfo{author}{Cheung, T.}, \bibinfo{author}{Schiavon, S.},
  \bibinfo{author}{Parkinson, T.}, \bibinfo{author}{Li, P.} \&
  \bibinfo{author}{Brager, G.}
\newblock \bibinfo{journal}{\bibinfo{title}{Analysis of the accuracy on
  pmv--ppd model using the ashrae global thermal comfort database ii}}.
\newblock {\emph{\JournalTitle{Building and Environment}}}
  \textbf{\bibinfo{volume}{153}}, \bibinfo{pages}{205--217}
  (\bibinfo{year}{2019}).

\bibitem{kim2018thermal}
\bibinfo{author}{Kim, J.} \& \bibinfo{author}{de~Dear, R.}
\newblock \bibinfo{journal}{\bibinfo{title}{Thermal comfort expectations and
  adaptive behavioural characteristics of primary and secondary school
  students}}.
\newblock {\emph{\JournalTitle{Building and Environment}}}
  \textbf{\bibinfo{volume}{127}}, \bibinfo{pages}{13--22}
  (\bibinfo{year}{2018}).

\bibitem{data_drivingstress}
\bibinfo{author}{Healey, J.~A.} \& \bibinfo{author}{Picard, R.~W.}
\newblock \bibinfo{journal}{\bibinfo{title}{Detecting stress during real-world
  driving tasks using physiological sensors}}.
\newblock {\emph{\JournalTitle{IEEE Transactions on Intelligent Transportation
  Systems}}} \textbf{\bibinfo{volume}{6}}, \bibinfo{pages}{156--166}
  (\bibinfo{year}{2005}).

\bibitem{data_deap}
\bibinfo{author}{Koelstra, S.} \emph{et~al.}
\newblock \bibinfo{journal}{\bibinfo{title}{Deap: A database for emotion
  analysis using physiological signals}}.
\newblock {\emph{\JournalTitle{IEEE Transactions on Affective Computing}}}
  \textbf{\bibinfo{volume}{3}}, \bibinfo{pages}{18--31} (\bibinfo{year}{2011}).

\bibitem{data_drivingwork}
\bibinfo{author}{Schneegass, S.}, \bibinfo{author}{Pfleging, B.},
  \bibinfo{author}{Broy, N.}, \bibinfo{author}{Heinrich, F.} \&
  \bibinfo{author}{Schmidt, A.}
\newblock \bibinfo{title}{A data set of real world driving to assess driver
  workload}.
\newblock In \emph{\bibinfo{booktitle}{Proceedings of the 5th International
  Conference on Automotive User Interfaces and Interactive Vehicular
  Applications}}, \bibinfo{pages}{150--157} (\bibinfo{year}{2013}).

\bibitem{data_studentlife}
\bibinfo{author}{Wang, R.} \emph{et~al.}
\newblock \bibinfo{title}{Studentlife: assessing mental health, academic
  performance and behavioral trends of college students using smartphones}.
\newblock In \emph{\bibinfo{booktitle}{Proceedings of the 2014 ACM
  International Joint Conference on Pervasive and Ubiquitous Computing}},
  \bibinfo{pages}{3--14} (\bibinfo{year}{2014}).

\bibitem{data_decaf}
\bibinfo{author}{Abadi, M.~K.} \emph{et~al.}
\newblock \bibinfo{journal}{\bibinfo{title}{Decaf: Meg-based multimodal
  database for decoding affective physiological responses}}.
\newblock {\emph{\JournalTitle{IEEE Transactions on Affective Computing}}}
  \textbf{\bibinfo{volume}{6}}, \bibinfo{pages}{209--222}
  (\bibinfo{year}{2015}).

\bibitem{data_noneeg}
\bibinfo{author}{Birjandtalab, J.}, \bibinfo{author}{Cogan, D.},
  \bibinfo{author}{Pouyan, M.~B.} \& \bibinfo{author}{Nourani, M.}
\newblock \bibinfo{title}{A non-eeg biosignals dataset for assessment and
  visualization of neurological status}.
\newblock In \emph{\bibinfo{booktitle}{2016 IEEE International Workshop on
  Signal Processing Systems (SiPS)}}, \bibinfo{pages}{110--114}
  (\bibinfo{organization}{IEEE}, \bibinfo{year}{2016}).

\bibitem{data_ascertain}
\bibinfo{author}{Subramanian, R.} \emph{et~al.}
\newblock \bibinfo{journal}{\bibinfo{title}{Ascertain: Emotion and personality
  recognition using commercial sensors}}.
\newblock {\emph{\JournalTitle{IEEE Transactions on Affective Computing}}}
  \textbf{\bibinfo{volume}{9}}, \bibinfo{pages}{147--160}
  (\bibinfo{year}{2016}).

\bibitem{data_stressmath}
\bibinfo{author}{Gjoreski, M.}, \bibinfo{author}{Lu{\v{s}}trek, M.},
  \bibinfo{author}{Gams, M.} \& \bibinfo{author}{Gjoreski, H.}
\newblock \bibinfo{journal}{\bibinfo{title}{Monitoring stress with a wrist
  device using context}}.
\newblock {\emph{\JournalTitle{Journal of Biomedical Informatics}}}
  \textbf{\bibinfo{volume}{73}}, \bibinfo{pages}{159--170}
  (\bibinfo{year}{2017}).

\bibitem{data_wesad}
\bibinfo{author}{Schmidt, P.}, \bibinfo{author}{Reiss, A.},
  \bibinfo{author}{Duerichen, R.}, \bibinfo{author}{Marberger, C.} \&
  \bibinfo{author}{Van~Laerhoven, K.}
\newblock \bibinfo{title}{Introducing wesad, a multimodal dataset for wearable
  stress and affect detection}.
\newblock In \emph{\bibinfo{booktitle}{Proceedings of the 20th ACM
  International Conference on Multimodal Interaction}},
  \bibinfo{pages}{400--408} (\bibinfo{year}{2018}).

\bibitem{data_cogload_snake}
\bibinfo{author}{Gjoreski, M.} \emph{et~al.}
\newblock \bibinfo{journal}{\bibinfo{title}{Datasets for cognitive load
  inference using wearable sensors and psychological traits}}.
\newblock {\emph{\JournalTitle{Applied Sciences}}}
  \textbf{\bibinfo{volume}{10}}, \bibinfo{pages}{3843} (\bibinfo{year}{2020}).

\bibitem{data_kemocon}
\bibinfo{author}{Park, C.~Y.} \emph{et~al.}
\newblock \bibinfo{journal}{\bibinfo{title}{K-emocon, a multimodal sensor
  dataset for continuous emotion recognition in naturalistic conversations}}.
\newblock {\emph{\JournalTitle{Scientific Data}}} \textbf{\bibinfo{volume}{7}},
  \bibinfo{pages}{1--16} (\bibinfo{year}{2020}).

\bibitem{gao2020n}
\bibinfo{author}{Gao, N.}, \bibinfo{author}{Shao, W.},
  \bibinfo{author}{Rahaman, M.~S.} \& \bibinfo{author}{Salim, F.~D.}
\newblock \bibinfo{journal}{\bibinfo{title}{n-gage: Predicting in-class
  emotional, behavioural and cognitive engagement in the wild}}.
\newblock {\emph{\JournalTitle{Proceedings of the ACM on Interactive, Mobile,
  Wearable and Ubiquitous Technologies}}} \textbf{\bibinfo{volume}{4}},
  \bibinfo{pages}{1--26} (\bibinfo{year}{2020}).

\bibitem{di2018unobtrusive}
\bibinfo{author}{Di~Lascio, E.}, \bibinfo{author}{Gashi, S.} \&
  \bibinfo{author}{Santini, S.}
\newblock \bibinfo{journal}{\bibinfo{title}{Unobtrusive assessment of students'
  emotional engagement during lectures using electrodermal activity sensors}}.
\newblock {\emph{\JournalTitle{Proceedings of the ACM on Interactive, Mobile,
  Wearable and Ubiquitous Technologies}}} \textbf{\bibinfo{volume}{2}},
  \bibinfo{pages}{1--21} (\bibinfo{year}{2018}).

\bibitem{bakker2011s}
\bibinfo{author}{Bakker, J.}, \bibinfo{author}{Pechenizkiy, M.} \&
  \bibinfo{author}{Sidorova, N.}
\newblock \bibinfo{title}{What's your current stress level? detection of stress
  patterns from gsr sensor data}.
\newblock In \emph{\bibinfo{booktitle}{2011 IEEE 11th International Conference
  on Data Mining Workshops}}, \bibinfo{pages}{573--580}
  (\bibinfo{organization}{IEEE}, \bibinfo{year}{2011}).

\bibitem{sarchiapone2018association}
\bibinfo{author}{Sarchiapone, M.} \emph{et~al.}
\newblock \bibinfo{journal}{\bibinfo{title}{The association between
  electrodermal activity (eda), depression and suicidal behaviour: A systematic
  review and narrative synthesis}}.
\newblock {\emph{\JournalTitle{BMC Sychiatry}}} \textbf{\bibinfo{volume}{18}},
  \bibinfo{pages}{1--27} (\bibinfo{year}{2018}).

\bibitem{handbook2009american}
\bibinfo{author}{Handbook-Fundamentals, A.}
\newblock \bibinfo{journal}{\bibinfo{title}{American society of heating}}.
\newblock {\emph{\JournalTitle{Refrigerating and Air-Conditioning Engineers}}}
  (\bibinfo{year}{2009}).

\bibitem{fuller2018development}
\bibinfo{author}{Fuller, K.~A.} \emph{et~al.}
\newblock \bibinfo{journal}{\bibinfo{title}{Development of a self-report
  instrument for measuring in-class student engagement reveals that pretending
  to engage is a significant unrecognized problem}}.
\newblock {\emph{\JournalTitle{PloS One}}} \textbf{\bibinfo{volume}{13}},
  \bibinfo{pages}{e0205828} (\bibinfo{year}{2018}).

\bibitem{pollak2011pam}
\bibinfo{author}{Pollak, J.~P.}, \bibinfo{author}{Adams, P.} \&
  \bibinfo{author}{Gay, G.}
\newblock \bibinfo{title}{Pam: a photographic affect meter for frequent, in
  situ measurement of affect}.
\newblock In \emph{\bibinfo{booktitle}{Proceedings of the SIGCHI Conference on
  Human Factors in Computing Systems}}, \bibinfo{pages}{725--734}
  (\bibinfo{year}{2011}).

\bibitem{garbarino2014empatica}
\bibinfo{author}{Garbarino, M.}, \bibinfo{author}{Lai, M.},
  \bibinfo{author}{Bender, D.}, \bibinfo{author}{Picard, R.~W.} \&
  \bibinfo{author}{Tognetti, S.}
\newblock \bibinfo{title}{Empatica e3—a wearable wireless multi-sensor device
  for real-time computerized biofeedback and data acquisition}.
\newblock In \emph{\bibinfo{booktitle}{2014 4th International Conference on
  Wireless Mobile Communication and Healthcare-Transforming Healthcare Through
  Innovations in Mobile and Wireless Technologies}}, \bibinfo{pages}{39--42}
  (\bibinfo{organization}{IEEE}, \bibinfo{year}{2014}).

\bibitem{gao2019predicting}
\bibinfo{author}{Gao, N.}, \bibinfo{author}{Shao, W.} \&
  \bibinfo{author}{Salim, F.~D.}
\newblock \bibinfo{journal}{\bibinfo{title}{Predicting personality traits from
  physical activity intensity}}.
\newblock {\emph{\JournalTitle{Computer}}} \textbf{\bibinfo{volume}{52}},
  \bibinfo{pages}{47--56} (\bibinfo{year}{2019}).

\bibitem{michael2020conversion}
\bibinfo{author}{Michael, P.~R.}, \bibinfo{author}{Johnston, D.~E.} \&
  \bibinfo{author}{Moreno, W.}
\newblock \bibinfo{journal}{\bibinfo{title}{A conversion guide: solar
  irradiance and lux illuminance}}.
\newblock {\emph{\JournalTitle{Journal of Measurements in Engineering}}}
  \textbf{\bibinfo{volume}{8}}, \bibinfo{pages}{153--166}
  (\bibinfo{year}{2020}).

\bibitem{wagner2018exploring}
\bibinfo{author}{Wagner, A.}, \bibinfo{author}{O’Brien, W.} \&
  \bibinfo{author}{Dong, B.}
\newblock \emph{\bibinfo{title}{Exploring occupant behavior in buildings:
  methods and challenges}} (\bibinfo{publisher}{Springer International
  Publishing}, \bibinfo{address}{Cham}, \bibinfo{year}{2018}).

\bibitem{gao_marschall_burry_watkins_salim_2021}
\bibinfo{author}{Gao, N.}, \bibinfo{author}{Marschall, M.},
  \bibinfo{author}{Burry, J.}, \bibinfo{author}{Watkins, S.} \&
  \bibinfo{author}{Salim, F.}
\newblock \bibinfo{title}{{In-Gauge} and {En-Gage} datasets}.
\newblock \bibinfo{howpublished}{\emph{Figshare}
  \url{https://doi.org/10.25439/rmt.14578908}} (\bibinfo{year}{2021}).

\bibitem{international2001international}
\bibinfo{author}{Taylor, B.~N.} \& \bibinfo{author}{Thompson, A.}
\newblock \emph{\bibinfo{title}{The international system of units (SI)}}
  (\bibinfo{publisher}{US Department of Commerce, Technology Administration,
  National Institute of Standards and Technology}, \bibinfo{year}{2001}).

\bibitem{braithwaite2013guide}
\bibinfo{author}{Braithwaite, J.~J.}, \bibinfo{author}{Watson, D.~G.},
  \bibinfo{author}{Jones, R.} \& \bibinfo{author}{Rowe, M.}
\newblock \bibinfo{journal}{\bibinfo{title}{A guide for analysing electrodermal
  activity (eda) \& skin conductance responses (scrs) for psychological
  experiments}}.
\newblock {\emph{\JournalTitle{Psychophysiology}}}
  \textbf{\bibinfo{volume}{49}}, \bibinfo{pages}{1017--1034}
  (\bibinfo{year}{2013}).

\bibitem{babaei2021critique}
\bibinfo{author}{Babaei, E.}, \bibinfo{author}{Tag, B.},
  \bibinfo{author}{Dingler, T.} \& \bibinfo{author}{Velloso, E.}
\newblock \bibinfo{title}{A critique of electrodermal activity practices at
  chi}.
\newblock In \emph{\bibinfo{booktitle}{Proceedings of the 2021 CHI Conference
  on Human Factors in Computing Systems}}, \bibinfo{pages}{1--14}
  (\bibinfo{year}{2021}).

\bibitem{gashi2020detection}
\bibinfo{author}{Gashi, S.} \emph{et~al.}
\newblock \bibinfo{journal}{\bibinfo{title}{Detection of artifacts in
  ambulatory electrodermal activity data}}.
\newblock {\emph{\JournalTitle{Proceedings of the ACM on Interactive, Mobile,
  Wearable and Ubiquitous Technologies}}} \textbf{\bibinfo{volume}{4}},
  \bibinfo{pages}{1--31} (\bibinfo{year}{2020}).

\bibitem{schwee2019room}
\bibinfo{author}{Schwee, J.~H.} \emph{et~al.}
\newblock \bibinfo{journal}{\bibinfo{title}{Room-level occupant counts and
  environmental quality from heterogeneous sensing modalities in a smart
  building}}.
\newblock {\emph{\JournalTitle{Scientific data}}} \textbf{\bibinfo{volume}{6}},
  \bibinfo{pages}{1--11} (\bibinfo{year}{2019}).

\bibitem{deldari2020espresso}
\bibinfo{author}{Deldari, S.}, \bibinfo{author}{Smith, D.~V.},
  \bibinfo{author}{Sadri, A.} \& \bibinfo{author}{Salim, F.}
\newblock \bibinfo{journal}{\bibinfo{title}{Espresso: Entropy and shape aware
  time-series segmentation for processing heterogeneous sensor data}}.
\newblock {\emph{\JournalTitle{Proceedings of the ACM on Interactive, Mobile,
  Wearable and Ubiquitous Technologies}}} \textbf{\bibinfo{volume}{4}},
  \bibinfo{pages}{1--24} (\bibinfo{year}{2020}).

\bibitem{shao2016clustering}
\bibinfo{author}{Shao, W.}, \bibinfo{author}{Salim, F.~D.},
  \bibinfo{author}{Song, A.} \& \bibinfo{author}{Bouguettaya, A.}
\newblock \bibinfo{journal}{\bibinfo{title}{Clustering big
  spatiotemporal-interval data}}.
\newblock {\emph{\JournalTitle{IEEE Transactions on Big Data}}}
  \textbf{\bibinfo{volume}{2}}, \bibinfo{pages}{190--203}
  (\bibinfo{year}{2016}).

\bibitem{shao2019onlineairtrajclus}
\bibinfo{author}{Shao, W.} \emph{et~al.}
\newblock \bibinfo{title}{Onlineairtrajclus: An online aircraft trajectory
  clustering for tarmac situation awareness}.
\newblock In \emph{\bibinfo{booktitle}{2019 IEEE International Conference on
  Pervasive Computing and Communications (PerCom}}, \bibinfo{pages}{192--201}
  (\bibinfo{organization}{IEEE}, \bibinfo{year}{2019}).

\bibitem{salim2020modelling}
\bibinfo{author}{Salim, F.~D.} \emph{et~al.}
\newblock \bibinfo{journal}{\bibinfo{title}{Modelling urban-scale occupant
  behaviour, mobility, and energy in buildings: A survey}}.
\newblock {\emph{\JournalTitle{Building and Environment}}}
  \textbf{\bibinfo{volume}{183}}, \bibinfo{pages}{106964}
  (\bibinfo{year}{2020}).

\bibitem{carlucci2020modeling}
\bibinfo{author}{Carlucci, S.} \emph{et~al.}
\newblock \bibinfo{journal}{\bibinfo{title}{Modeling occupant behavior in
  buildings}}.
\newblock {\emph{\JournalTitle{Building and Environment}}}
  \textbf{\bibinfo{volume}{174}}, \bibinfo{pages}{106768}
  (\bibinfo{year}{2020}).

\bibitem{kjaergaard2020current}
\bibinfo{author}{Kj{\ae}rgaard, M.~B.} \emph{et~al.}
\newblock \bibinfo{journal}{\bibinfo{title}{Current practices and
  infrastructure for open data based research on occupant-centric design and
  operation of buildings}}.
\newblock {\emph{\JournalTitle{Building and Environment}}}
  \textbf{\bibinfo{volume}{177}}, \bibinfo{pages}{106848}
  (\bibinfo{year}{2020}).

\bibitem{rahaman2020ambient}
\bibinfo{author}{Rahaman, M.~S.} \emph{et~al.}
\newblock \bibinfo{journal}{\bibinfo{title}{An ambient--physical system to
  infer concentration in open-plan workplace}}.
\newblock {\emph{\JournalTitle{IEEE Internet of Things Journal}}}
  \textbf{\bibinfo{volume}{7}}, \bibinfo{pages}{11576--11586}
  (\bibinfo{year}{2020}).

\bibitem{gashi2018using}
\bibinfo{author}{Gashi, S.}, \bibinfo{author}{Di~Lascio, E.} \&
  \bibinfo{author}{Santini, S.}
\newblock \bibinfo{title}{Using students' physiological synchrony to quantify
  the classroom emotional climate}.
\newblock In \emph{\bibinfo{booktitle}{Proceedings of the 2018 ACM
  International Joint Conference and 2018 International Symposium on Pervasive
  and Ubiquitous Computing and Wearable Computers}}, \bibinfo{pages}{698--701}
  (\bibinfo{year}{2018}).

\bibitem{sacerdote2014experimental}
\bibinfo{author}{Sacerdote, B.}
\newblock \bibinfo{journal}{\bibinfo{title}{Experimental and quasi-experimental
  analysis of peer effects: two steps forward?}}
\newblock {\emph{\JournalTitle{Annu. Rev. Econ.}}}
  \textbf{\bibinfo{volume}{6}}, \bibinfo{pages}{253--272}
  (\bibinfo{year}{2014}).

\bibitem{gao2021investigating}
\bibinfo{author}{Gao, N.}, \bibinfo{author}{Saiedur~Rahaman, M.},
  \bibinfo{author}{Shao, W.} \& \bibinfo{author}{Salim, F.~D.}
\newblock \bibinfo{title}{Investigating the reliability of self-report data in
  the wild: The quest for ground truth}.
\newblock In \emph{\bibinfo{booktitle}{Adjunct Proceedings of the 2021 ACM
  International Joint Conference on Pervasive and Ubiquitous Computing and
  Proceedings of the 2021 ACM International Symposium on Wearable Computers}},
  \bibinfo{pages}{237--242} (\bibinfo{year}{2021}).

\bibitem{park2020heat}
\bibinfo{author}{Park, R.~J.}, \bibinfo{author}{Goodman, J.},
  \bibinfo{author}{Hurwitz, M.} \& \bibinfo{author}{Smith, J.}
\newblock \bibinfo{journal}{\bibinfo{title}{Heat and learning}}.
\newblock {\emph{\JournalTitle{American Economic Journal: Economic Policy}}}
  \textbf{\bibinfo{volume}{12}}, \bibinfo{pages}{306--39}
  (\bibinfo{year}{2020}).

\bibitem{arief2017hoc}
\bibinfo{author}{Arief-Ang, I.~B.}, \bibinfo{author}{Salim, F.~D.} \&
  \bibinfo{author}{Hamilton, M.}
\newblock \bibinfo{title}{Da-hoc: semi-supervised domain adaptation for room
  occupancy prediction using co2 sensor data}.
\newblock In \emph{\bibinfo{booktitle}{Proceedings of the 4th ACM International
  Conference on Systems for Energy-Efficient Built Environments}},
  \bibinfo{pages}{1--10} (\bibinfo{year}{2017}).

\bibitem{arief2018rup}
\bibinfo{author}{Arief-Ang, I.~B.}, \bibinfo{author}{Hamilton, M.} \&
  \bibinfo{author}{Salim, F.~D.}
\newblock \bibinfo{journal}{\bibinfo{title}{Rup: Large room utilisation
  prediction with carbon dioxide sensor}}.
\newblock {\emph{\JournalTitle{Pervasive and Mobile Computing}}}
  \textbf{\bibinfo{volume}{46}}, \bibinfo{pages}{49--72}
  (\bibinfo{year}{2018}).

\bibitem{arief2018scalable}
\bibinfo{author}{Arief-Ang, I.~B.}, \bibinfo{author}{Hamilton, M.} \&
  \bibinfo{author}{Salim, F.~D.}
\newblock \bibinfo{journal}{\bibinfo{title}{A scalable room occupancy
  prediction with transferable time series decomposition of co2 sensor data}}.
\newblock {\emph{\JournalTitle{ACM Transactions on Sensor Networks (TOSN)}}}
  \textbf{\bibinfo{volume}{14}}, \bibinfo{pages}{1--28} (\bibinfo{year}{2018}).

\bibitem{mccarthy2016validation}
\bibinfo{author}{McCarthy, C.}, \bibinfo{author}{Pradhan, N.},
  \bibinfo{author}{Redpath, C.} \& \bibinfo{author}{Adler, A.}
\newblock \bibinfo{title}{Validation of the empatica e4 wristband}.
\newblock In \emph{\bibinfo{booktitle}{2016 IEEE EMBS international student
  conference (ISC)}}, \bibinfo{pages}{1--4} (\bibinfo{organization}{IEEE},
  \bibinfo{year}{2016}).

\bibitem{menghini2019stressing}
\bibinfo{author}{Menghini, L.} \emph{et~al.}
\newblock \bibinfo{journal}{\bibinfo{title}{Stressing the accuracy: Wrist-worn
  wearable sensor validation over different conditions}}.
\newblock {\emph{\JournalTitle{Psychophysiology}}}
  \textbf{\bibinfo{volume}{56}}, \bibinfo{pages}{e13441}
  (\bibinfo{year}{2019}).

\end{thebibliography}

\section*{Acknowledgements} 
This research is supported by the Australian Government through the Australian Research Council’s Linkage Projects funding scheme (project LP150100246 in partnership with Aurecon). This paper is also a contribution to the IEA EBC Annex 79.

\section*{Author contributions statement}
N.G. designed, prepared, and conducted the cross-sectional study for wearable data collection, analysed the wearable and indoor sensor data, and wrote the manuscript. M.M. designed, prepared and conducted the longitudinal data collection for outdoor and indoor sensing, and processed the dataset and wrote the manuscript. 
J.B, S.W., and F.D.S  supervised the data collection, dataset design, and revised the manuscript. F.D.S advised N.G on the overall project and data analysis and modelling. 

\section*{Competing interests}
The authors declare no competing interests.

\end{document}